\documentclass[pra,showpacs,superscriptaddress]{revtex4}
\usepackage{amsmath}
\usepackage{amsfonts}
\usepackage{amssymb}
\usepackage{graphicx}
\usepackage{bbm}
\usepackage{natbib}
\usepackage[dvipdfm]{hyperref}
\makeatletter
\begin{document}
\title{Realization of hierarchical equations of motion from stochastic perspectives}
\author{Wei Wu}

\email{weiwu@csrc.ac.cn}

\affiliation{Beijing Computational Science Research Center, Beijing 100193, People's Republic of China}

\begin{abstract}
The hierarchical equations of motion (HEOM) for a generalized quantum dissipative system is rigorously constructed in the frameworks of two different stochastic dynamical descriptions, i.e., the non-Markovian quantum state diffusion approach as well as the stochastic decoupling scheme. We demonstrated that the HEOMs obtained by these two different stochastic dynamical methods are identical. Moreover, we present some numerical examples to verify the feasibility of our formalism.
\end{abstract}
\maketitle

\section{Introduction}\label{sec:sec1}

The rapid development of nanotechnology has opened the possibility to realize some physical, chemical and biological tasks at an atomic scale in current available experiments~\cite{1}. Many microcosmic systems, such as the trapped ion~\cite{2,3}, the nuclear spin~\cite{4,5,6} and the superconducting circuit~\cite{7,8,9}, are promising candidates for these tasks which have been demonstrated in many previous studies. On the other hand, due to the unavoidable coupling with the surrounding bath, the microscopic quantum device severely undergoes decoherence which is the main difficulty in fulfilling reliable quantum experiments. In this sense, an accurate description of the dynamics of a quantum system embedded in a bath is urgently required. Despite a long history of theoretical studies, the dynamical problem of a quantum dissipative system remains a challenging problem. Up to now, to the best of our knowledge, there is still no reliable approach to simulating the dynamics of an arbitrary quantum dissipative system without approximations.

In many examples on the literature, dissipation-induced decoherence or relaxation in a quantum microscopic system is modeled by the spin-boson model, which describes the interaction between a spin subsystem and a set of non-interacting harmonic oscillators. The spin-boson model has attracted considerable attention in past decades because it provides a very simple model to simulate many practical processes. For instance, the spin-boson model and its extensions have been used to investigate the optical spectroscopy of molecular aggregates~\cite{10,11} and the electron energy transfer dynamics in the Fenna-Matthews-Olson complex~\cite{12,13}. The reduced system dynamics of the spin-boson model has been studied by various analytical and numerical methods, such as, the generalized Silbey-Harris transformation approach~\cite{14,15,16}, the time-dependent numerical renormalization-group technique~\cite{17,18}, the Dirac-Frenkel time-dependent variation scheme~\cite{19}, the non-Markovian quantum state diffusion (NMQSD) method~\cite{20,21,22} and the stochastic decoupling (SD) formalism~\cite{23,24,25}. Each method has its own regimes of validity depending on the system-bath coupling strength, the bath temperature, and the bath spectral density function. The NMQSD approach, which was presented by Strunz, Yu, and their coworkers, constitutes a theory for a non-Markovian quantum trajectory that is applicable in both the bosonic bath and the fermionic bath situations. The main thought of the NMQSD is to establish a connection between the coherent state description of the bath and the stochastic interpretation of the equation of motion. Shao and co-workers put forward a stochastic Liouville equation to describe the dynamical behaviour of a quantum dissipative system by decoupling the system-bath interaction with the help of the Hubbard-Stratonovich transformation or the characteristics of the It$\hat{\mathrm{o}}$ calculus. The basic idea of the SD scheme developed by Shao \emph{et al}. is to convert the effect of the system-bath correlation into a random field exerting on the evolution of the quantum subsystem.

NMQSD and SD are two different stochastic dynamical methods. In these stochastic schemes, the quantum subsystem is subjected to random noises which characterize the environmental influences on the quantum subsystem. The main advantage of the stochastic dynamical description is its conceptual simplicity: the sole effect of the bath is providing a stochastic field which randomizes the equation of motion of the quantum subsystem~\cite{20,21,22,23,24,25,26,27}. However, within the stochastic dynamical scheme, one has to handle numerous random quantum trajectories in order to reach a convergent statistical expectation which makes the stochastic simulation rather time-consuming. Moreover, a stochastic dynamical method is usually adopted to simulate the short-time dynamical behaviour, while in the long-time limit, the stochastic simulation becomes unstable. Many previous studies tried to improved the controllability of the stochastic simulation in the long-time regime by adding nonlinearity in the equation of motion~\cite{22,28}. However, we believe that a more natural way to eliminate this problem is to convert the stochastic description to a deterministic quantum master equation by taking the average over all the realizations of the noises~\cite{26,27,29,30}, because numerically solving a deterministic quantum master equation is usually much easier than performing a stochastic dynamical simulation.

However, it is rather ideal that a deterministic quantum master equation can be extracted from a stochastic description, for many quantum dissipative systems, no closed quantum master equation can be achieved. Fortunately, a deterministic HEOM can be extracted within the framework of stochastic dynamical schemes. This hybrid method, which combines the merits of the stochastic and the deterministic schemes, has become a hot topic in quantum dissipative dynamics~\cite{26,27,29,30,add0}. The HEOM is a set of time-local differential equations for the reduced density matrix of the quantum subsystem, which was originally proposed by Tanimura and co-workers~\cite{31,32,33}. This numerical treatment includes all the orders of the system-bath interactions, and it is beyond the usual Markovian approximation, the rotating-wave approximation, and the perturbative approximation. The HEOM can be viewed as a bridge linking the original Schrodinger equation or quantum Liouville equation of a quantum open system, which is usually difficult to handle, and a set of ordinary differential equations, which can be numerically solved by using the Runge-Kutta method. How to establish such a bridge, which should be elaborately designed without losing any important dynamical information of a quantum open system, is the crucial step of the HEOM treatment. In many previous studies, the HEOM was realized by employing the path integral influence functional approach~\cite{34,35,36,37,38} which is rather complicated in mathematics. In this paper, we adopt an alternative way to establish the HEOM and believe that any new viewpoint of describing the dissipative dynamics of a quantum open system would be helpful to obtaining more physical insights into this research field. Moreover, we also rigorously demonstrate that the HEOMs obtained by making use of the NMQSD and the SD methods are identical. This is a very interesting result, because, usually, the expressions of the HEOMs derived by different approaches are more or less different. Considering the fact that the HEOM provides numerically exact simulations of quantum dissipative dynamics, our result suggests that the NMQSD and the SD methods may be equivalent in terms of the description of dissipative dynamics.

In this paper, we consider a generalized Hamiltonian of a quantum dissipative system which can be described by
\begin{equation}\label{eq:eq1}
\hat{H}=\hat{H}_{\mathrm{s}}+\hat{H}_{\mathrm{b}}+\hat{S}^{\dag}\hat{B}+\hat{S}\hat{B}^{\dag},
\end{equation}
where $\hat{H}_{\mathrm{s}}$ is the quantum subsystem of interest, and the operator $\hat{S}$ denotes the quantum subsystem's dissipative operator, which couples to the surrounding environment. The Hamiltonian of the bath is given by $\hat{H}_{\mathrm{b}}\equiv\sum_{\ell}\omega_{\ell}\hat{b}_{\ell}^{\dag}\hat{b}_{\ell}$, where $\hat{b}_{\ell}$ and $\hat{b}_{\ell}^{\dagger}$ are the bosonic annihilation and creation operators of the $\ell$-th environmental mode with frequency $\omega_{\ell}$, respectively. Operator $\hat{B}\equiv\sum_{\ell}g_{\ell}^{*}\hat{b}_{\ell}$ is the bath's dissipative operator, where parameters $g_{\ell}$ are complex numbers quantifying the coupling strength between the quantum subsystem and its environment. In many previous studies of the SD approach~\cite{39,40,add1,add2}, the authors restricted their attentions to the case $\hat{S}=\hat{S}^{\dag}$, our formulation is beyond this limitation. In this paper, we assume that the bath is initially prepared in the thermal equilibrium state, $\hat{\varrho}_{\mathrm{th}}\equiv e^{-\beta \hat{H}_{\mathrm{b}}}/\mathrm{tr}_{\mathrm{b}}(e^{-\beta \hat{H}_{\mathrm{b}}})$.

The paper is organized as follows: In Sec.~\ref{sec:sec2}, the realization of the HEOMs by making use of the NMQSD method and the SD scheme are rigorously presented. In Sec.~\ref{sec:sec3}, we make some comparisons between the numerical results from the HEOM and some well-known results obtained by other approaches. Some discussions and the main conclusions of this paper are presented in Sec.~\ref{sec:sec4}. In several Appendixes, we provide some additional details about the main text. Throughout the paper, we set $\hbar=k_{\mathrm{B}}=1$, and all the other units are dimensionless as well.

\section{Theory}\label{sec:sec2}

In this section, we shall show how to rigorously derive the HEOMs by making use of two different stochastic dynamical formulations. In the derivation, complex-valued stochastic processes are involved for both the NMQSD method and the SD technique. To avoid misunderstanding, we use different notations to distinguish the stochastic processes involved in the NMQSD and the SD approaches, respectively. And considering the fact that the SD method is rooted in the It$\hat{\mathrm{o}}$ formalism, we prefer to use the term, ``Wiener process", which is usually associated with the It$\hat{\mathrm{o}}$ calculus, to identify the stochastic process involved in the SD scheme. In Appendix~\ref{sec:secappa}, we provide additional materials on the properties of these stochastic processes to clarify some details of the present work.

\subsection{NMQSD approach}\label{sec:sec2a}

The NMQSD method was originally developed for zero-temperature bosonic bath systems. This approach requires us to assume that the bath is prepared in its Fock vacuum state~\cite{20,21,22}. However, the finite-temperature NMQSD method can be easily generalized from the zero-temperature case by employing the so-called thermofield formulation which maps a thermal equilibrium state onto a zero-temperature vacuum state by adding degrees of freedom~\cite{41,42}. To be specific, for the total Hamiltonian given by Eq.~\ref{eq:eq1}, we introduce a fictitious bath whose Hamiltonian is given by $\hat{H}_{\mathrm{f}}\equiv\sum_{\ell}(-\omega_{\ell})\hat{f}_{\ell}^{\dag}\hat{f}_{\ell}$, here $\hat{f}_{\ell}^{\dag}$ and $\hat{f}_{\ell}$ are independent fictitious bosonic creation and annihilation operators, respectively. One can find that the $\ell$-th mode of the fictitious bath has the same but negative frequency $(-\omega_{\ell})$ as that for the physical bath. As a consequence, the composite bath, i.e., the physical bath plus the fictitious bath, is given by
\begin{equation*}
\hat{H}_{\mathrm{bf}}=\sum_{\ell}\omega_{\ell}(\hat{b}_{\ell}^{\dag}\hat{b}_{\ell}-\hat{f}_{\ell}^{\dag}\hat{f}_{\ell}),
\end{equation*}
which doubles the degrees of freedom compared with that of the physical bath. The desired zero-temperature vacuum state $|\mathrm{\mathbf{vac}}\rangle$ of the composite bath is now constructed such that one can recover the correct thermal equilibrium state $\hat{\varrho}_{\mathrm{th}}$ of the physical bath after tracing out the fictitious degrees of freedom, i.e.,
\begin{equation*}
\hat{\varrho}_{\mathrm{th}}=\mathrm{tr}_{\mathrm{f}}(|\mathrm{\mathbf{vac}}\rangle\langle\mathrm{\mathbf{vac}}|).
\end{equation*}
The explicit expression of $|\mathrm{\mathbf{vac}}\rangle$ will be determined later. It is necessary to point out that the additional degrees of freedom do not alter the dissipative dynamics, because they are uncoupled to the physical ones~\cite{41,42}. The main purpose of introducing these additional degrees of freedom is to purifying the thermal equilibrium state $\hat{\varrho}_{\mathrm{th}}$. By doing so, all the formulations which are originally established in the zero-temperature case, can be easily extended to the finite-temperature situation without difficulties.

The corresponding new Hamiltonian of the total system then reads
\begin{equation*}
\hat{H}_{\beta}=\hat{H}_{\mathrm{s}}+\sum_{\ell}\omega_{\ell}\Big{(}\hat{b}_{\ell}^{\dag}\hat{b}_{\ell}-\hat{f}_{\ell}^{\dag}\hat{f}_{\ell}\Big{)}+\sum_{\ell}\Big{(}g_{\ell}\hat{S}^{\dag}\hat{b}_{\ell}+g_{\ell}^{*}\hat{S}\hat{b}_{\ell}^{\dagger}\Big{)}.
\end{equation*}
We apply a temperature-dependent Bogoliubov transformation to the above Hamiltonian
\begin{equation*}
\hat{b}_{\ell}\equiv\sqrt{n(\omega_{\ell})+1}\hat{c}_{\ell}+\sqrt{n(\omega_{\ell})}\hat{d}_{\ell}^{\dag},
\end{equation*}
\begin{equation*}
\hat{f}_{\ell}\equiv\sqrt{n(\omega_{\ell})+1}\hat{d}_{\ell}+\sqrt{n(\omega_{\ell})}\hat{c}_{\ell}^{\dag},
\end{equation*}
where $n(\omega_{\ell})$ is the mean thermal occupation number, i.e., $n(\omega_{\ell})\equiv (e^{\beta\omega_{\ell}}-1)^{-1}$. The transformed Hamiltonian is then given by
\begin{equation*}
\hat{H}_{\beta}=\hat{H}_{\mathrm{s}}+\sum_{\ell}\sqrt{n(\omega_{\ell})+1}\Big{(}g_{\ell}\hat{S}\hat{c}_{\ell}^{\dagger}+g_{\ell}^{*}\hat{S}^{\dagger}\hat{c}_{\ell}\Big{)}+\sum_{\ell}\omega_{\ell}\hat{c}_{\ell}^{\dag}\hat{c}_{\ell}+\sum_{\ell}\sqrt{n(\omega_{\ell})}\Big{(}g_{\ell}\hat{S}\hat{d}_{\ell}+g_{\ell}^{*}\hat{S}^{\dagger}\hat{d}_{\ell}^{\dagger}\Big{)}-\sum_{\ell}\omega_{\ell}\hat{d}_{\ell}^{\dag}\hat{d}_{\ell}.
\end{equation*}
Then, the expression of the desired zero-temperature vacuum state of the composite bath is given by $|\mathrm{\mathbf{vac}}\rangle\equiv\prod_{\ell}|\mathrm{vac}_{\ell}\rangle\otimes|\mathrm{\widetilde{vac}}_{\ell}\rangle$, satisfying $\hat{c}_{\ell}|\mathrm{vac}_{\ell}\rangle=0$ and $\hat{d}_{\ell}|\mathrm{\widetilde{vac}}_{\ell}\rangle=0$.

The dynamics of the new Hamiltonian of the whole system, i.e., $\hat{H}_{\beta}$, is governed by the Schr$\ddot{\mathrm{o}}$dinger equation, $\partial_{t}|\Psi(t)\rangle=-i\hat{H}_{\beta}|\Psi(t)\rangle$ with initial state $|\Psi(0)\rangle=|\Phi(0)\rangle\otimes|\mathrm{\mathbf{vac}}\rangle$, where $|\Psi(t)\rangle$ is the pure-state wave function of the whole system, and $|\Phi(0)\rangle$ denotes the initial state of the quantum subsystem. Practically, due to the large number of degrees of freedom in the environment, it is impossible to exactly solve this Schr$\ddot{\mathrm{o}}$dinger equation. However, by introducing the Bargmann coherent state~\cite{43} $|\mathbf{z}\mathbf{\tilde{z}}\rangle\equiv|\mathbf{z}\rangle\otimes|\mathbf{\tilde{z}}\rangle$, where $|\textbf{z}\rangle\equiv\prod_{\ell}\exp(z_{\ell}\hat{c}_{\ell}^{\dagger})|\mathrm{vac}_{\ell}\rangle$ and $|\mathbf{\tilde{z}}\rangle\equiv\prod_{\ell}\exp({\tilde{z}}_{\ell}\hat{d}_{\ell}^{\dagger})|\mathrm{\widetilde{vac}}_{\ell}\rangle$, one can recast the Schr$\ddot{\mathrm{o}}$dinger equation into a stochastic Schr$\ddot{\mathrm{o}}$dinger equation of the quantum state diffusion type~\cite{41,42}
\begin{equation}\label{eq:eq2}
\begin{split}
\partial_{t}|\Psi(\mathbf{z}_{t}^{*},\mathbf{\tilde{z}}_{t}^{*})\rangle=&-i\hat{H}_{\mathrm{s}}|\Psi(\mathbf{z}_{t}^{*},\mathbf{\tilde{z}}_{t}^{*})\rangle+\hat{S}\mathbf{z}_{t}^{*}|\Psi(\mathbf{z}_{t}^{*},\mathbf{\tilde{z}}_{t}^{*})\rangle+\hat{S}^{\dag}\mathbf{\tilde{z}}_{t}^{*}|\Psi(\mathbf{z}_{t}^{*},\mathbf{\tilde{z}}_{t}^{*})\rangle\\
&-\hat{S}^{\dag}\int_{0}^{t}d\tau\alpha(t-\tau)\frac{\delta}{\delta \mathbf{z}_{\tau}^{*}}|\Psi(\mathbf{z}_{t}^{*},\mathbf{\tilde{z}}_{t}^{*})\rangle-\hat{S}\int_{0}^{t}d\tau\tilde{\alpha}(t-\tau)\frac{\delta}{\delta \mathbf{\tilde{z}}_{\tau}^{*}}|\Psi(\mathbf{z}_{t}^{*},\mathbf{\tilde{z}}_{t}^{*})\rangle,
\end{split}
\end{equation}
where $|\Psi(\mathbf{\mathbf{z}}_{t}^{*},\mathbf{\tilde{z}}_{t}^{*})\rangle\equiv\langle \mathbf{z}\mathbf{\tilde{z}}|\Psi(t)\rangle$ is the total pure-state wave function under the Bargmann coherent state representation. The variables $\mathbf{z}_{t}\equiv i\sum_{\ell}\sqrt{n(\omega_{\ell})+1}g_{\ell}z_{\ell}e^{-i\omega_{\ell}t}$ and $\tilde{\mathbf{z}}_{t}\equiv i\sum_{\ell}\sqrt{n(\omega_{\ell})}g_{\ell}\tilde{z}_{\ell}e^{i\omega_{\ell}t}$ can be interpreted as two independent Gaussian processes, $\alpha(t)\equiv\sum_{\ell}|g_{\ell}|^{2}[n(\omega_{\ell})+1]e^{-i\omega_{\ell}t}$ and $\tilde{\alpha}(t)\equiv\sum_{\ell}|g_{\ell}|^{2}n(\omega_{\ell})e^{i\omega_{\ell}t}$ are two modified bath correlation functions. By introducing the so-called bath density spectral function $J(\omega)\equiv\sum_{\ell}|g_{\ell}|^{2}\delta(\omega-\omega_{\ell})$, they can be rewritten as follows
\begin{equation}\label{eq:eq3}
\alpha(t)=\int_{0}^{\infty} d\omega J(\omega)[n(\omega)+1]e^{-i\omega t},
\end{equation}
\begin{equation}\label{eq:eq4}
\tilde{\alpha}(t)=\int_{0}^{\infty} d\omega J(\omega)n(\omega)e^{i\omega t}.
\end{equation}
It is easy to check that $\tilde{\alpha}(t)$ vanishes in the limit $\beta\rightarrow \infty$, while, $\alpha(t)$ reduces to the zero-temperature bath correlation function in Refs.~\cite{20,21,22} in the same limitation.

For the sake convenience, we reexpress these two modified bath correlation functions as $\alpha(t)=\alpha_{\mathrm{R}}(t)+i\alpha_{\mathrm{I}}(t)$ and $\tilde{\alpha}(t)=\tilde{\alpha}_{\mathrm{R}}(t)+i\tilde{\alpha}_{\mathrm{I}}(t)$, where the subscripts $\mathrm{R}$ and $\mathrm{I}$ stand for the real and imaginary parts, respectively. To realize the HEOM numerical performance, it is important that the modified bath correlation functions can be (or at least approximately) written as finite sums of exponentials:
\begin{equation}\label{eq:eq5}
\alpha_{\mathrm{X}}(t)\simeq\sum_{n_{\mathrm{X}}=1}^{N_{\mathrm{X}}}U_{n_{\mathrm{X}}}^{\mathrm{X}}\exp(-V_{n_{\mathrm{X}}}^{\mathrm{X}}t),
\end{equation}
\begin{equation}\label{eq:eq6}
\tilde{\alpha}_{\mathrm{X}}(t)\simeq\sum_{\tilde{n}_{\mathrm{X}}=1}^{\tilde{N}_{\mathrm{X}}}\tilde{U}_{\tilde{n}_{\mathrm{X}}}^{\mathrm{X}}\exp(-\tilde{V}_{\tilde{n}_{\mathrm{X}}}^{\mathrm{X}}t),
\end{equation}
where the fitting numbers $\{U_{n_{\mathrm{X}}}^{\mathrm{X}},V_{n_{\mathrm{X}}}^{\mathrm{X}},N_{\mathrm{X}},\tilde{U}_{\tilde{n}_{\mathrm{X}}}^{\mathrm{X}},\tilde{V}_{\tilde{n}_{\mathrm{X}}}^{\mathrm{X}},\tilde{N}_{\mathrm{X}}\}$ are allowed to be uncorrelated for the real $\mathrm{X}=\mathrm{R}$ and the imaginary $\mathrm{X}=\mathrm{I}$ parts. It is necessary to point out that our fitting given by Eq.~\ref{eq:eq5} and Eq.~\ref{eq:eq6} is different from that of Refs.~\cite{28,29}. In Refs.~\cite{28,29}, the authors expand the whole bath correlation function (real and imaginary parts) as a sum of exponentials. For a practical numerical fitting, we believe that it is more convenient to fit the real and the imaginary parts of the bath correlation function, respectively. Except for some special examples, an exact decomposition of the modified bath correlation functions cannot be achieved. Nevertheless, many different approaches have been proposed to try to obtain a good fit of the bath correlation function with as few exponential terms as possible~\cite{44,45,46,47}, because, typically, the numerical cost of performing the HEOM scheme grows rapidly with the number of exponentials in Eq.~\ref{eq:eq5} and Eq.~\ref{eq:eq6}. In Appendix~\ref{sec:secappb}, we briefly show how to find the good fitting functions for these two modified bath correlation functions by making use of the so-called sum-over-poles decomposition metho, which was commonly adopted in many previous studies~\cite{44,45,46,47}.

While Eq.~\ref{eq:eq2} determines the dynamics of the whole system exactly, in general, it is still unclear how to solve Eq.~\ref{eq:eq2} due to the functional derivatives. In many previous works~\cite{20,21,22,41}, the functional derivatives were replaced by new operators acting in the quantum subsystem's Hilbert space, i.e.,
\begin{equation*}
\frac{\delta}{\delta\mathbf{z}_{\tau}^{*}}|\Psi(\mathbf{z}_{t}^{*},\mathbf{\tilde{z}}_{t}^{*})\rangle=\mathcal{\hat{O}}(t,\tau,\mathbf{z}_{t}^{*},\mathbf{\tilde{z}}_{t}^{*})|\Psi(\mathbf{z}_{t}^{*},\mathbf{\tilde{z}}_{t}^{*})\rangle, \end{equation*}
and
\begin{equation*} \frac{\delta}{\delta\mathbf{\tilde{z}}_{\tau}^{*}}|\Psi(\mathbf{z}_{t}^{*},\mathbf{\tilde{z}}_{t}^{*})\rangle=\mathcal{\hat{G}}(t,\tau,\mathbf{z}_{t}^{*},\mathbf{\tilde{z}}_{t}^{*})|\Psi(\mathbf{z}_{t}^{*},\mathbf{\tilde{z}}_{t}^{*})\rangle. \end{equation*}
For some special cases~\cite{48,49}, the new operators $\mathcal{\hat{O}}(t,\tau,\mathbf{z}_{t}^{*},\mathbf{\tilde{z}}_{t}^{*})$ and $\mathcal{\hat{G}}(t,\tau,\mathbf{z}_{t}^{*},\mathbf{\tilde{z}}_{t}^{*})$ can be determined exactly. However, in general, these new operators are unknown. In this paper, we use the hierarchy of the pure-state wave functions method, which was originally reported in Refs.~\cite{28,29}, to avoid directly handle these functional derivatives and corresponding memory integrals. The procedure is outlined as follows:

First, let us define the following four super-operators
\begin{equation}\label{eq:eq7}
\mathfrak{\hat{D}}_{n_{\mathrm{X}}}^{\mathrm{X}}=(i)^{\theta_{\mathrm{X}}}\int_{0}^{t}d\tau U_{n_{\mathrm{X}}}^{\mathrm{X}}\exp[-V_{n_{\mathrm{X}}}^{\mathrm{X}}(t-\tau)]\frac{\delta}{\delta \mathbf{z}_{\tau}^{*}},
\end{equation}
\begin{equation}\label{eq:eq8}
\hat{\mathfrak{E}}_{\tilde{n}_{\mathrm{X}}}^{\mathrm{X}}=(i)^{\theta_{\mathrm{X}}}\int_{0}^{t}d\tau \tilde{U}_{\tilde{n}_{\mathrm{X}}}^{\mathrm{X}}\exp[-\tilde{V}_{\tilde{n}_{\mathrm{X}}}^{\mathrm{X}}(t-\tau)]\frac{\delta}{\delta \mathbf{\tilde{z}}_{\tau}^{*}},
\end{equation}
where $\theta_{\mathrm{R}}=0$ and $\theta_{\mathrm{I}}=1$. Considering the fact that the fitting parameters can be complex, the pre-factor $(i)^{\theta_{\mathrm{X}}}$ can be absorbed into the choice of $U_{n_{\mathrm{X}}}^{\mathrm{X}}$ and $\tilde{U}_{\tilde{n}_{\mathrm{X}}}^{\mathrm{X}}$. For the sake of convenience, we shall also define the general expression of the auxiliary pure-state wave functions as follows
\begin{equation}\label{eq:eq9}
|\Psi_{\mathbf{p},\mathbf{q},\tilde{\mathbf{p}},\tilde{\mathbf{q}}}\rangle=|\Psi_{\mathbf{p}_{N_{\mathrm{R}}},\mathbf{q}_{N_{\mathrm{I}}},\tilde{\mathbf{p}}_{\tilde{N}_{\mathrm{R}}},\tilde{\mathbf{q}}_{\tilde{N}_{\mathrm{I}}}}\rangle\equiv\Big{[}(\mathfrak{\hat{D}}_{n_{\mathrm{R}}}^{\mathrm{R}})^{\mathrm{P}_{n_{\mathrm{R}}}}(\mathfrak{\hat{D}}_{n_{\mathrm{I}}}^{\mathrm{I}})^{\mathrm{Q}_{n_{\mathrm{I}}}}(\hat{\mathfrak{E}}_{\tilde{n}_{\mathrm{R}}}^{\mathrm{R}})^{\tilde{\mathrm{P}}_{\tilde{n}_{\mathrm{R}}}}(\hat{\mathfrak{E}}_{\tilde{n}_{\mathrm{I}}}^{\mathrm{I}})^{\tilde{\mathrm{Q}}_{\tilde{n}_{\mathrm{I}}}}\Big{]}|\Psi(\mathbf{z}_{t}^{*},\tilde{\mathbf{z}}_{t}^{*})\rangle,
\end{equation}
here the vector $\mathbf{a}=\mathbf{p},\mathbf{q},\tilde{\mathbf{p}},\tilde{\mathbf{q}}$ is defined as $\mathbf{a}=\mathbf{a}_{\mathrm{M}}\equiv\{A_{\mathrm{m}}\}_{1}^{\mathrm{M}}=\{A_{\mathrm{m}}\}_{\mathrm{m}=1}^{\mathrm{m}=\mathrm{M}}=\{A_{1},A_{2},A_{3},...,A_{\mathrm{m}},...,A_{\mathrm{M}}\}$, where the subscript $\mathrm{M}$ stands for its dimension and $A_{\mathrm{m}}$ indicates its $\mathrm{m}$-th element. Then, Eq.~\ref{eq:eq2} can be rewritten as
\begin{equation}\label{eq:eq10}
\begin{split}
\partial_{t}|\Psi_{\mathbf{0},\mathbf{0},\mathbf{0},\mathbf{0}}\rangle=&(-i\hat{H}_{\mathrm{s}}+\hat{S}\mathbf{z}_{t}^{*}+\hat{S}^{\dag}\tilde{\mathbf{z}}_{t}^{*})|\Psi_{\mathbf{0},\mathbf{0},\mathbf{0},\mathbf{0}}\rangle\\
&-\hat{S}^{\dag}\sum_{n_{\mathrm{R}}}|\Psi_{0...1_{n_{\mathrm{R}}}...0,\mathbf{0},\mathbf{0},\mathbf{0}}\rangle-\hat{S}^{\dag}\sum_{n_{\mathrm{I}}}|\Psi_{\mathbf{0},0...1_{n_{\mathrm{I}}}...0,\mathbf{0},\mathbf{0}}\rangle\\
&-\hat{S}\sum_{\tilde{n}_{\mathrm{R}}}|\Psi_{\mathbf{0},\mathbf{0},0...1_{\tilde{n}_{\mathrm{R}}}...0,\mathbf{0}}\rangle-\hat{S}\sum_{\tilde{n}_{\mathrm{I}}}|\Psi_{\mathbf{0},\mathbf{0},\mathbf{0},0...1_{\tilde{n}_{\mathrm{I}}}...0}\rangle,
\end{split}
\end{equation}
where we omitted the upper-scripts in all the sum-notations for the sake of consentience. In this paper, we refer the sum of $n_{\mathrm{X}}(\tilde{n}_{\mathrm{X}})$ is from $1$ to $N_{\mathrm{X}}(\tilde{N}_{\mathrm{X}})$ without an explicit statement. In the expression of Eq.~\ref{eq:eq9}, we have defined the following very useful notations to replace the functional derivatives and corresponding memory integrals
\begin{equation*}
\mathfrak{\hat{D}}_{1}^{\mathrm{R}}|\Psi(\mathbf{z}_{t}^{*},\mathbf{\tilde{z}}_{t}^{*})\rangle=|\Psi_{100...0,\mathbf{0},\mathbf{0},\mathbf{0}}\rangle,~\mathfrak{\hat{D}}_{2}^{\mathrm{R}}|\Psi(\mathbf{z}_{t}^{*},\mathbf{\tilde{z}}_{t}^{*})\rangle=|\Psi_{010...0,\mathbf{0},\mathbf{0},\mathbf{0}}\rangle,~\mathfrak{\hat{D}}_{3}^{\mathrm{R}}|\Psi(\mathbf{z}_{t}^{*},\mathbf{\tilde{z}}_{t}^{*})\rangle=|\Psi_{001...0,\mathbf{0},\mathbf{0},\mathbf{0}}\rangle,...
\end{equation*}
\begin{equation*}
\mathfrak{\hat{D}}_{1}^{\mathrm{I}}|\Psi(\mathbf{z}_{t}^{*},\mathbf{\tilde{z}}_{t}^{*})\rangle=|\Psi_{\mathbf{0},100...0,\mathbf{0},\mathbf{0}}\rangle,~\mathfrak{\hat{D}}_{2}^{\mathrm{I}}|\Psi(\mathbf{z}_{t}^{*},\mathbf{\tilde{z}}_{t}^{*})\rangle=|\Psi_{\mathbf{0},010...0,\mathbf{0},\mathbf{0}}\rangle,~\mathfrak{\hat{D}}_{3}^{\mathrm{I}}|\Psi(\mathbf{z}_{t}^{*},\mathbf{\tilde{z}}_{t}^{*})\rangle=|\Psi_{\mathbf{0},001...0,\mathbf{0},\mathbf{0}}\rangle,...
\end{equation*}
\begin{equation*}
\mathfrak{\hat{E}}_{1}^{\mathrm{R}}|\Psi(\mathbf{z}_{t}^{*},\mathbf{\tilde{z}}_{t}^{*})\rangle=|\Psi_{\mathbf{0},\mathbf{0},100...0,\mathbf{0}}\rangle,~\mathfrak{\hat{E}}_{2}^{\mathrm{R}}|\Psi(\mathbf{z}_{t}^{*},\mathbf{\tilde{z}}_{t}^{*})\rangle=|\Psi_{\mathbf{0},\mathbf{0},010...0,\mathbf{0}}\rangle,~\mathfrak{\hat{E}}_{3}^{\mathrm{R}}|\Psi(\mathbf{z}_{t}^{*},\mathbf{\tilde{z}}_{t}^{*})\rangle=|\Psi_{\mathbf{0},\mathbf{0},001...0,\mathbf{0}}\rangle,...
\end{equation*}
\begin{equation*}
\mathfrak{\hat{E}}_{1}^{\mathrm{I}}|\Psi(\mathbf{z}_{t}^{*},\mathbf{\tilde{z}}_{t}^{*})\rangle=|\Psi_{\mathbf{0},\mathbf{0},\mathbf{0},100...0}\rangle,~\mathfrak{\hat{E}}_{2}^{\mathrm{I}}|\Psi(\mathbf{z}_{t}^{*},\mathbf{\tilde{z}}_{t}^{*})\rangle=|\Psi_{\mathbf{0},\mathbf{0},\mathbf{0},010...0}\rangle,~\mathfrak{\hat{E}}_{3}^{\mathrm{I}}|\Psi(\mathbf{z}_{t}^{*},\mathbf{\tilde{z}}_{t}^{*})\rangle=|\Psi_{\mathbf{0},\mathbf{0},\mathbf{0},001...0}\rangle,....
\end{equation*}
As you can see from Eq.~\ref{eq:eq9}, the solution of $|\Psi(\mathbf{z}_{t}^{*},\mathbf{\tilde{z}}_{t}^{*})\rangle=|\Psi_{\mathbf{0},\mathbf{0},\mathbf{0},\mathbf{0}}\rangle$ is determined by its own free evolution as well as the dynamics of auxiliary pure-state wave functions in the second and third rows of Eq.~\ref{eq:eq9}. To solve these auxiliary pure-state wave functions, we need their differential equations. Next, we would like to show how to derive the dynamics of $|\Psi_{0...1_{n_{\mathrm{R}}}...0,\mathbf{0},\mathbf{0},\mathbf{0}}\rangle$. One can obtain the time evolution of all the other auxiliary pure-state wave functions by using the same procedure.

The bounded integral domain in Eq.~\ref{eq:eq7} and Eq.~\ref{eq:eq8} can be extended from $[0,t]$ to $[-\infty,+\infty]$ due to the fact that the pure-state wave function should be independent of the noise-terms $\mathbf{z}_{s}$ and $\mathbf{\tilde{z}}_{s}$ for $s<0$ and $s>t$~\cite{28}. Thus, one can find that
\begin{equation*}
\partial_{t}\mathfrak{\hat{D}}_{n_{\mathrm{X}}}^{\mathrm{X}}=-V_{n_{\mathrm{X}}}^{\mathrm{X}}\mathfrak{\hat{D}}_{n_{\mathrm{X}}}^{\mathrm{X}};~~\partial_{t}\mathfrak{\hat{E}}_{\tilde{n}_{\mathrm{X}}}^{\mathrm{X}}=-\tilde{V}_{\tilde{n}_{\mathrm{X}}}^{\mathrm{X}}\mathfrak{\hat{E}}_{\tilde{n}_{\mathrm{X}}}^{\mathrm{X}}.
\end{equation*}
Then, taking the time derivative of $|\Psi_{0...1_{n_{\mathrm{R}}}...0,\mathbf{0},\mathbf{0},\mathbf{0}}\rangle$, one can obtain
\begin{equation}\label{eq:eq11}
\begin{split}
\partial_{t}|\Psi_{0...1_{n_{\mathrm{R}}}...0,\mathbf{0},\mathbf{0},\mathbf{0}}\rangle=&\partial_{t}[\mathfrak{\hat{D}}_{n_{\mathrm{R}}}^{\mathrm{R}}|\Psi_{\mathbf{0},\mathbf{0},\mathbf{0},\mathbf{0}}\rangle]\\
=&(\partial_{t}\mathfrak{\hat{D}}_{n_{\mathrm{R}}}^{\mathrm{R}})|\Psi_{\mathbf{0},\mathbf{0},\mathbf{0},\mathbf{0}}\rangle+\mathfrak{\hat{D}}_{n_{\mathrm{R}}}^{\mathrm{R}}\partial_{t}|\Psi_{\mathbf{0},\mathbf{0},\mathbf{0},\mathbf{0}}\rangle\\
=&-V_{n_{\mathrm{R}}}^{\mathrm{R}}\mathfrak{\hat{D}}_{n_{\mathrm{R}}}^{\mathrm{R}}|\Psi_{\mathbf{0},\mathbf{0},\mathbf{0},\mathbf{0}}\rangle-i\hat{H}_{\mathrm{s}}\mathfrak{\hat{D}}_{n_{\mathrm{R}}}^{\mathrm{R}}|\Psi_{\mathbf{0},\mathbf{0},\mathbf{0},\mathbf{0}}\rangle+\hat{S}\mathfrak{\hat{D}}_{n_{\mathrm{R}}}^{\mathrm{R}}\mathbf{z}_{t}^{*}|\Psi_{\mathbf{0},\mathbf{0},\mathbf{0},\mathbf{0}}\rangle+\hat{S}^{\dag}\mathbf{\tilde{z}}_{t}^{*}\mathfrak{\hat{D}}_{n_{\mathrm{R}}}^{\mathrm{R}}|\Psi_{\mathbf{0},\mathbf{0},\mathbf{0},\mathbf{0}}\rangle\\
&-\hat{S}^{\dag}\sum_{n_{\mathrm{R}}'}\mathfrak{\hat{D}}_{n_{\mathrm{R}}}^{\mathrm{R}}|\Psi_{0...1_{n_{\mathrm{R}}'}...0,\mathbf{0},\mathbf{0},\mathbf{0}}\rangle-\hat{S}^{\dag}\sum_{n_{\mathrm{I}}}\mathfrak{\hat{D}}_{n_{\mathrm{R}}}^{\mathrm{R}}|\Psi_{\mathbf{0},0...1_{n_{\mathrm{I}}}...0,\mathbf{0},\mathbf{0}}\rangle\\
&-\hat{S}\sum_{\tilde{n}_{R}}\mathfrak{\hat{D}}_{n_{\mathrm{R}}}^{\mathrm{R}}|\Psi_{\mathbf{0},\mathbf{0},0...1_{\tilde{n}_{\mathrm{R}}},\mathbf{0}}\rangle-\hat{S}\sum_{\tilde{n}_{\mathrm{I}}}\mathfrak{\hat{D}}_{n_{\mathrm{R}}}^{\mathrm{R}}|\Psi_{\mathbf{0},\mathbf{0},\mathbf{0},0...1_{\tilde{n}_{\mathrm{I}}}...0}\rangle\\
=&(-i\hat{H}_{\mathrm{s}}-V_{n_{\mathrm{R}}}^{\mathrm{R}}+\hat{S}\mathbf{z}_{t}^{*}+\hat{S}^{\dag}\mathbf{\tilde{z}}_{t}^{*})|\Psi_{0...1_{n_{\mathrm{R}}}...0,\mathbf{0},\mathbf{0},\mathbf{0}}\rangle+U_{n_{\mathrm{R}}}^{\mathrm{R}}\hat{S}|\Psi_{\mathbf{0},\mathbf{0},\mathbf{0},\mathbf{0}}\rangle\\
&-\hat{S}^{\dag}\sum_{n_{\mathrm{R}}'}|\Psi_{0...1_{n_{\mathrm{R}}}...1_{n_{\mathrm{R}}'}...0,\mathbf{0},\mathbf{0},\mathbf{0}}\rangle-\hat{S}^{\dag}\sum_{n_{\mathrm{I}}}|\Psi_{0...1_{n_{\mathrm{R}}}...0,0...1_{n_{\mathrm{I}}}...0,\mathbf{0},\mathbf{0}}\rangle\\
&-\hat{S}\sum_{\tilde{n}_{R}}|\Psi_{0...1_{n_{\mathrm{R}}}...0,\mathbf{0},0...1_{\tilde{n}_{\mathrm{R}}},\mathbf{0}}\rangle-\hat{S}\sum_{\tilde{n}_{\mathrm{I}}}|\Psi_{0...1_{n_{\mathrm{R}}}...0,\mathbf{0},\mathbf{0},0...1_{\tilde{n}_{\mathrm{I}}}...0}\rangle,
\end{split}
\end{equation}
where we have used the commutator relation $[\mathfrak{\hat{D}}_{n_{\mathrm{R}}}^{\mathrm{R}},\mathbf{z}_{t}^{*}]=U_{n_{\mathrm{R}}}^{\mathrm{R}}$ in the derivation. As you can see from Eq.~\ref{eq:eq11}, the differential equation of $|\Psi_{0...1_{n_{\mathrm{R}}}...0,\mathbf{0},\mathbf{0},\mathbf{0}}\rangle$ is coupled to more auxiliary pure-state wave functions. By repeating the same procedure shown in the derivation of $\partial_{t}|\Psi_{0...1_{n_{\mathrm{R}}}...0,\mathbf{0},\mathbf{0},\mathbf{0}}\rangle$ case, one can finally obtain the hierarchy equation of pure-state wave functions,
\begin{equation}\label{eq:eq12}
\begin{split}
\partial_{t}|\Psi_{\mathbf{p},\mathbf{q},\tilde{\mathbf{p}},\tilde{\mathbf{q}}}\rangle=&(-i\hat{H}_{\mathrm{s}}-\mathbf{p}\cdot \mathbf{v_{R}}-\mathbf{q}\cdot \mathbf{v_{I}}-\mathbf{\tilde{p}}\cdot\mathbf{\tilde{v}_{R}}-\mathbf{\tilde{q}}\cdot\mathbf{\tilde{v}_{I}}+\hat{S}\mathbf{z}_{t}^{*}+\hat{S}^{\dag}\tilde{\mathbf{z}}_{t}^{*})|\Psi_{\mathbf{p},\mathbf{q},\tilde{\mathbf{p}},\tilde{\mathbf{q}}}\rangle+\sum_{n_{\mathrm{R}}}P_{n_{\mathrm{R}}}U_{n_{\mathrm{R}}}^{\mathrm{R}}\hat{S}|\Psi_{\mathbf{p}-\mathbf{e}_{n_{\mathrm{R}}},\mathbf{q},\tilde{\mathbf{p}},\tilde{\mathbf{q}}}\rangle\\
&+\sum_{n_{\mathrm{I}}}Q_{n_{\mathrm{I}}}U_{n_{\mathrm{I}}}^{\mathrm{I}}\hat{S}|\Psi_{\mathbf{p},\mathbf{q}-\mathbf{e}_{n_{\mathrm{I}}},\tilde{\mathbf{p}},\tilde{\mathbf{q}}}\rangle+\sum_{\tilde{n}_{\mathrm{R}}}\tilde{P}_{\tilde{n}_{\mathrm{R}}}\tilde{U}_{\tilde{n}_{\mathrm{R}}}^{\mathrm{R}}\hat{S}^{\dag}|\Psi_{\mathbf{p},\mathbf{q},\tilde{\mathbf{p}}-\mathbf{e}_{\tilde{n}_{\mathrm{R}}},\tilde{\mathbf{q}}}\rangle+\sum_{\tilde{n}_{\mathrm{I}}}\tilde{Q}_{\tilde{n}_{\mathrm{I}}}\tilde{U}_{n_{\mathrm{I}}}^{\mathrm{I}}\hat{S}^{\dagger}|\Psi_{\mathbf{p},\mathbf{q},\tilde{\mathbf{p}},\tilde{\mathbf{q}}-\mathbf{e}_{\tilde{n}_{\mathrm{I}}}}\rangle\\
&-\hat{S}^{\dag}\Big{(}\sum_{n_{\mathrm{R}}}|\Psi_{\mathbf{p}+\mathbf{e}_{n_{\mathrm{R}}},\mathbf{q},\tilde{\mathbf{p}},\tilde{\mathbf{q}}}\rangle+\sum_{n_{\mathrm{I}}}|\Psi_{\mathbf{p},\mathbf{q}+\mathbf{e}_{n_{\mathrm{I}}},\tilde{\mathbf{p}},\tilde{\mathbf{q}}}\rangle\Big{)}-\hat{S}\Big{(}\sum_{\tilde{n}_{\mathrm{R}}}|\Psi_{\mathbf{p},\mathbf{q},\tilde{\mathbf{p}}+\mathbf{e}_{\tilde{n}_{\mathrm{R}}},\tilde{\mathbf{q}}}\rangle+\sum_{\tilde{n}_{\mathrm{I}}}|\Psi_{\mathbf{p},\mathbf{q},\tilde{\mathbf{p}},\tilde{\mathbf{q}}+\mathbf{e}_{\tilde{n}_{\mathrm{I}}}}\rangle\Big{)},
\end{split}
\end{equation}
where $\mathbf{v_{X}}\equiv\mathbf{v}_{N_{\mathrm{X}}}^{\mathrm{X}}=\{V_{1}^{\mathrm{X}},V_{2}^{\mathrm{X}},V_{3}^{\mathrm{X}},...,V_{N_{\mathrm{X}}}^{\mathrm{X}}\}$, $\mathbf{\tilde{v}_{X}}\equiv\mathbf{\tilde{v}}_{\tilde{N}_{\mathrm{X}}}^{\mathrm{X}}=\{\tilde{V}_{1}^{\mathrm{X}},\tilde{V}_{2}^{\mathrm{X}},\tilde{V}_{3}^{\mathrm{X}},...,\tilde{V}_{\tilde{N}_{\mathrm{X}}}^{\mathrm{X}}\}$ and $\mathbf{e}_{j}=\{0,0,0,...,1_{j},...,0\}$ is a $\check{N}$-dimensional vector, where $\check{N}\equiv\max\{N_{\mathrm{X}},\tilde{N}_{\mathrm{X}}\}$.

The hierarchy equation of pure-state wave functions in Eq.~\ref{eq:eq12} no longer contains the functional derivative, however, Eq.~\ref{eq:eq12} still contains the stochastic Gaussian noises, which hinders the efficiency of a numerical simulation. To remove these stochastic noises, one needs to take the statistical mean over all the possible stochastic processes which is equivalent to tracing out the degrees of freedom of the environment~\cite{20,21,22,28,29,41,48}. Then the hierarchy equation of the reduced density matrix is given by
\begin{equation*}
\begin{split}
\partial_{t}\hat{\varrho}_{\mathbf{p},\mathbf{q},\mathbf{\tilde{p}},\mathbf{\tilde{q}}}^{\mathbf{k},\mathbf{l},\mathbf{\tilde{k}},\mathbf{\tilde{l}}}=&\frac{\partial}{\partial t}\mathbb{M}\{|\Psi_{\mathbf{p},\mathbf{q},\tilde{\mathbf{p}},\tilde{\mathbf{q}}}\rangle\langle\Psi_{\mathbf{k},\mathbf{l},\tilde{\mathbf{k}},\tilde{\mathbf{l}}}|\}\\
=&\mathbb{M}\Bigg{\{}\frac{\overrightarrow{\partial}}{\partial t}|\Psi_{\mathbf{p},\mathbf{q},\tilde{\mathbf{p}},\tilde{\mathbf{q}}}\rangle\langle\Psi_{\mathbf{k},\mathbf{l},\tilde{\mathbf{k}},\tilde{\mathbf{l}}}|\Bigg{\}}+\mathbb{M}\Bigg{\{}|\Psi_{\mathbf{p},\mathbf{q},\tilde{\mathbf{p}},\tilde{\mathbf{q}}}\rangle\langle\Psi_{\mathbf{k},\mathbf{l},\tilde{\mathbf{k}},\tilde{\mathbf{l}}}|\frac{\overleftarrow{\partial}}{\partial t}\Bigg{\}},
\end{split}
\end{equation*}
where $\overleftarrow{\partial}$ and $\overrightarrow{\partial}$ are the left and right time derivative with respect to $|\Psi_{\mathbf{p},\mathbf{q},\tilde{\mathbf{p}},\tilde{\mathbf{q}}}\rangle$, respectively. With the help of Eq.~\ref{eq:eq12}, one can immediately find that
\begin{equation*}
\begin{split}
\partial_{t}\hat{\varrho}_{\mathbf{p},\mathbf{q},\mathbf{\tilde{p}},\mathbf{\tilde{q}}}^{\mathbf{k},\mathbf{l},\mathbf{\tilde{k}},\mathbf{\tilde{l}}}=&(-i\hat{H}_{\mathrm{s}}-\mathbf{p}\cdot \mathbf{v_{R}}-\mathbf{q}\cdot \mathbf{v_{I}}-\mathbf{\tilde{p}}\cdot\mathbf{\tilde{v}_{R}}-\mathbf{\tilde{q}}\cdot\mathbf{\tilde{v}_{I}})\hat{\varrho}_{\mathbf{p},\mathbf{q},\mathbf{\tilde{p}},\mathbf{\tilde{q}}}^{\mathbf{k},\mathbf{l},\mathbf{\tilde{k}},\mathbf{\tilde{l}}}+\hat{S}\mathbb{M}\{\mathbf{z}_{t}^{*}|\Psi_{\mathbf{p},\mathbf{q},\tilde{\mathbf{p}},\tilde{\mathbf{q}}}\rangle\langle\Psi_{\mathbf{k},\mathbf{l},\tilde{\mathbf{k}},\tilde{\mathbf{l}}}|\}+\hat{S}^{\dag}\mathbb{M}\{\mathbf{\tilde{z}}_{t}^{*}|\Psi_{\mathbf{p},\mathbf{q},\tilde{\mathbf{p}},\tilde{\mathbf{q}}}\rangle\langle\Psi_{\mathbf{k},\mathbf{l},\tilde{\mathbf{k}},\tilde{\mathbf{l}}}|\}\\
+&\sum_{n_{\mathrm{R}}}P_{n_{\mathrm{R}}}U_{n_{\mathrm{R}}}^{\mathrm{R}}\hat{S}\hat{\varrho}_{\mathbf{p}-\mathbf{e}_{n_{\mathrm{R}}},\mathbf{q},\mathbf{\tilde{p}},\mathbf{\tilde{q}}}^{\mathbf{k},\mathbf{l},\mathbf{\tilde{k}},\mathbf{\tilde{l}}}+\sum_{n_{\mathrm{I}}}Q_{n_{\mathrm{I}}}U_{n_{\mathrm{I}}}^{\mathrm{I}}\hat{S}\hat{\varrho}_{\mathbf{p},\mathbf{q}-\mathbf{e}_{n_{\mathrm{I}}},\mathbf{\tilde{p}},\mathbf{\tilde{q}}}^{\mathbf{k},\mathbf{l},\mathbf{\tilde{k}},\mathbf{\tilde{l}}}+\sum_{\tilde{n}_{\mathrm{R}}}\tilde{P}_{\tilde{n}_{\mathrm{R}}}\tilde{U}_{\tilde{n}_{\mathrm{R}}}^{\mathrm{R}}\hat{S}^{\dag}\hat{\varrho}_{\mathbf{p},\mathbf{q},\mathbf{\tilde{p}}-\mathbf{e}_{\tilde{n}_{\mathrm{R}}},\mathbf{\tilde{q}}}^{\mathbf{k},\mathbf{l},\mathbf{\tilde{k}},\mathbf{\tilde{l}}}+\sum_{\tilde{n}_{\mathrm{I}}}\tilde{Q}_{\tilde{n}_{\mathrm{I}}}\tilde{U}_{\tilde{n}_{\mathrm{I}}}^{\mathrm{I}}\hat{S}^{\dagger}\hat{\varrho}_{\mathbf{p},\mathbf{q},\mathbf{\tilde{p}},\mathbf{\tilde{q}}-\mathbf{e}_{\tilde{n}_{\mathrm{I}}}}^{\mathbf{k},\mathbf{l},\mathbf{\tilde{k}},\mathbf{\tilde{l}}}\\
-&\hat{S}^{\dag}\sum_{n_{\mathrm{R}}}\hat{\varrho}_{\mathbf{p}+\mathbf{e}_{n_{\mathrm{R}}},\mathbf{q},\mathbf{\tilde{p}},\mathbf{\tilde{q}}}^{\mathbf{k},\mathbf{l},\mathbf{\tilde{k}},\mathbf{\tilde{l}}}-\hat{S}^{\dag}\sum_{n_{\mathrm{I}}}\hat{\varrho}_{\mathbf{p},\mathbf{q}+\mathbf{e}_{n_{\mathrm{I}}},\mathbf{\tilde{p}},\mathbf{\tilde{q}}}^{\mathbf{k},\mathbf{l},\mathbf{\tilde{k}},\mathbf{\tilde{l}}}-\hat{S}\sum_{\tilde{n}_{\mathrm{R}}}\hat{\varrho}_{\mathbf{p},\mathbf{q},\mathbf{\tilde{p}}+\mathbf{e}_{\tilde{n}_{\mathrm{R}}},\mathbf{\tilde{q}}}^{\mathbf{k},\mathbf{l},\mathbf{\tilde{k}},\mathbf{\tilde{l}}}-\hat{S}\sum_{\tilde{n}_{\mathrm{I}}}\hat{\varrho}_{\mathbf{p},\mathbf{q},\mathbf{\tilde{p}},\mathbf{\tilde{q}}+\mathbf{e}_{\tilde{n}_{\mathrm{I}}}}^{\mathbf{k},\mathbf{l},\mathbf{\tilde{k}},\mathbf{\tilde{l}}}\\
+&\hat{\varrho}_{\mathbf{p},\mathbf{q},\mathbf{\tilde{p}},\mathbf{\tilde{q}}}^{\mathbf{k},\mathbf{l},\mathbf{\tilde{k}},\mathbf{\tilde{l}}}(i\hat{H}_{\mathrm{s}}-\mathbf{k}\cdot \mathbf{v_{R}^{*}}-\mathbf{l}\cdot \mathbf{v_{I}^{*}}-\mathbf{\tilde{k}}\cdot\mathbf{\tilde{v}_{R}^{*}}-\mathbf{\tilde{l}}\cdot\mathbf{\tilde{v}_{I}^{*}})+\mathbb{M}\{|\Psi_{\mathbf{p},\mathbf{q},\tilde{\mathbf{p}},\tilde{\mathbf{q}}}\rangle\langle\Psi_{\mathbf{k},\mathbf{l},\tilde{\mathbf{k}},\tilde{\mathbf{l}}}|\mathbf{z}_{t}\}\hat{S}^{\dagger}+\mathbb{M}\{|\Psi_{\mathbf{p},\mathbf{q},\tilde{\mathbf{p}},\tilde{\mathbf{q}}}\rangle\langle\Psi_{\mathbf{k},\mathbf{l},\tilde{\mathbf{k}},\tilde{\mathbf{l}}}|\mathbf{\tilde{z}}_{t}\}\hat{S}\\
+&\sum_{n_{\mathrm{R}}}K_{n_{\mathrm{R}}}U_{n_{\mathrm{R}}}^{\mathrm{R}*}\hat{\varrho}_{\mathbf{p},\mathbf{q},\mathbf{\tilde{p}},\mathbf{\tilde{q}}}^{\mathbf{k}-\mathbf{e}_{n_{\mathrm{R}}},\mathbf{l},\mathbf{\tilde{k}},\mathbf{\tilde{l}}}\hat{S}^{\dagger}+\sum_{n_{\mathrm{I}}}L_{n_{\mathrm{I}}}U_{n_{\mathrm{I}}}^{\mathrm{I}*}\hat{\varrho}_{\mathbf{p},\mathbf{q},\mathbf{\tilde{p}},\mathbf{\tilde{q}}}^{\mathbf{k},\mathbf{l}-\mathbf{e}_{n_{\mathrm{I}}},\mathbf{\tilde{k}},\mathbf{\tilde{l}}}\hat{S}^{\dagger}+\sum_{\tilde{n}_{\mathrm{R}}}\tilde{K}_{\tilde{n}_{\mathrm{R}}}\tilde{U}_{\tilde{n}_{\mathrm{R}}}^{\mathrm{R}*}\hat{\varrho}_{\mathbf{p},\mathbf{q},\mathbf{\tilde{p}},\mathbf{\tilde{q}}}^{\mathbf{k},\mathbf{l},\mathbf{\tilde{k}}-\mathbf{e}_{\tilde{n}_{\mathrm{R}}},\mathbf{\tilde{l}}}\hat{S}+\sum_{\tilde{n}_{\mathrm{I}}}\tilde{L}_{\tilde{n}_{\mathrm{I}}}\tilde{U}_{\tilde{n}_{\mathrm{I}}}^{\mathrm{I}*}\hat{\varrho}_{\mathbf{p},\mathbf{q},\mathbf{\tilde{p}},\mathbf{\tilde{q}}}^{\mathbf{k},\mathbf{l},\mathbf{\tilde{k}},\mathbf{\tilde{l}}-\mathbf{e}_{\tilde{n}_{\mathrm{I}}}}\hat{S}\\
-&\sum_{n_{\mathrm{R}}}\hat{\varrho}_{\mathbf{p},\mathbf{q},\mathbf{\tilde{p}},\mathbf{\tilde{q}}}^{\mathbf{k}+\mathbf{e}_{n_{\mathrm{R}}},\mathbf{l},\mathbf{\tilde{k}},\mathbf{\tilde{l}}}\hat{S}-\sum_{n_{\mathrm{I}}}\hat{\varrho}_{\mathbf{p},\mathbf{q},\mathbf{\tilde{p}},\mathbf{\tilde{q}}}^{\mathbf{k},\mathbf{l}+\mathbf{e}_{n_{\mathrm{I}}},\mathbf{\tilde{k}},\mathbf{\tilde{l}}}\hat{S}-\sum_{\tilde{n}_{\mathrm{R}}}\hat{\varrho}_{\mathbf{p},\mathbf{q},\mathbf{\tilde{p}},\mathbf{\tilde{q}}}^{\mathbf{k},\mathbf{l},\mathbf{\tilde{k}}+\mathbf{e}_{\tilde{n}_{\mathrm{R}}},\mathbf{\tilde{l}}}\hat{S}^{\dagger}-\sum_{\tilde{n}_{\mathrm{I}}}\hat{\varrho}_{\mathbf{p},\mathbf{q},\mathbf{\tilde{p}},\mathbf{\tilde{q}}}^{\mathbf{k},\mathbf{l},\mathbf{\tilde{k}},\mathbf{\tilde{l}}+\mathbf{e}_{\tilde{n}_{\mathrm{I}}}}\hat{S}^{\dagger},
\end{split}
\end{equation*}
where $\mathbf{v_{X}^{*}}$ and $\mathbf{\tilde{v}_{X}^{*}}$ denote the complex conjugate of all the elements in vectors $\mathbf{v_{X}}$ and $\mathbf{\tilde{v}_{X}}$, respectively. The above equation still contains some stochastic Gaussian-noise terms in the averages, however, these noise terms can be eliminated by making use of the extended Furutsu-Novikov theorem~\cite{22,29,50}. As we show in Appendix~\ref{sec:secappc}, one can demonstrate that
\begin{equation}\label{eq:eq13}
\mathbb{M}\{|\Psi_{\mathbf{p},\mathbf{q},\tilde{\mathbf{p}},\tilde{\mathbf{q}}}\rangle\langle\Psi_{\mathbf{k},\mathbf{l},\tilde{\mathbf{k}},\tilde{\mathbf{l}}}|\mathbf{z}_{t}\}=\sum_{n_{\mathrm{R}}}\hat{\varrho}_{\mathbf{p}+\mathbf{e}_{n_{\mathrm{R}}},\mathbf{q},\mathbf{\tilde{p}},\mathbf{\tilde{q}}}^{\mathbf{k},\mathbf{l},\mathbf{\tilde{k}},\mathbf{\tilde{l}}}+\sum_{n_{\mathrm{I}}}\hat{\varrho}_{\mathbf{p},\mathbf{q}+\mathbf{e}_{n_{\mathrm{I}}},\mathbf{\tilde{p}},\mathbf{\tilde{q}}}^{\mathbf{k},\mathbf{l},\mathbf{\tilde{k}},\mathbf{\tilde{l}}}.
\end{equation}
\begin{equation}\label{eq:eq14}
\mathbb{M}\{|\Psi_{\mathbf{p},\mathbf{q},\tilde{\mathbf{p}},\tilde{\mathbf{q}}}\rangle\langle\Psi_{\mathbf{k},\mathbf{l},\tilde{\mathbf{k}},\tilde{\mathbf{l}}}|\mathbf{\tilde{z}}_{t}\}=\sum_{\tilde{n}_{\mathrm{R}}}\hat{\varrho}_{\mathbf{p},\mathbf{q},\mathbf{\tilde{p}}+\mathbf{e}_{\tilde{n}_{\mathrm{R}}},\mathbf{\tilde{q}}}^{\mathbf{k},\mathbf{l},\mathbf{\tilde{k}},\mathbf{\tilde{l}}}+\sum_{\tilde{n}_{\mathrm{I}}}\hat{\varrho}_{\mathbf{p},\mathbf{q},\mathbf{\tilde{p}},\mathbf{\tilde{q}}+\mathbf{e}_{\tilde{n}_{\mathrm{I}}}}^{\mathbf{k},\mathbf{l},\mathbf{\tilde{k}},\mathbf{\tilde{l}}}.
\end{equation}
\begin{equation}\label{eq:eq15}
\mathbb{M}\{\mathbf{z}_{t}^{*}|\Psi_{\mathbf{p},\mathbf{q},\tilde{\mathbf{p}},\tilde{\mathbf{q}}}\rangle\langle\Psi_{\mathbf{k},\mathbf{l},\tilde{\mathbf{k}},\tilde{\mathbf{l}}}|\}=\sum_{n_{\mathrm{R}}}\hat{\varrho}_{\mathbf{p},\mathbf{q},\mathbf{\tilde{p}},\mathbf{\tilde{q}}}^{\mathbf{k}+\mathbf{e}_{n_{\mathrm{R}}},\mathbf{l},\mathbf{\tilde{k}},\mathbf{\tilde{l}}}+\sum_{n_{\mathrm{I}}}\hat{\varrho}_{\mathbf{p},\mathbf{q},\mathbf{\tilde{p}},\mathbf{\tilde{q}}}^{\mathbf{k},\mathbf{l}+\mathbf{e}_{n_{\mathrm{I}}},\mathbf{\tilde{k}},\mathbf{\tilde{l}}}.
\end{equation}
\begin{equation}\label{eq:eq16}
\mathbb{M}\{\mathbf{\tilde{z}}_{t}^{*}|\Psi_{\mathbf{p},\mathbf{q},\tilde{\mathbf{p}},\tilde{\mathbf{q}}}\rangle\langle\Psi_{\mathbf{k},\mathbf{l},\tilde{\mathbf{k}},\tilde{\mathbf{l}}}|\}=\sum_{\tilde{n}_{\mathrm{R}}}\hat{\varrho}_{\mathbf{p},\mathbf{q},\mathbf{\tilde{p}},\mathbf{\tilde{q}}}^{\mathbf{k},\mathbf{l},\mathbf{\tilde{k}}+\mathbf{e}_{\tilde{n}_{\mathrm{R}}},\mathbf{\tilde{l}}}+\sum_{\tilde{n}_{\mathrm{I}}}\hat{\varrho}_{\mathbf{p},\mathbf{q},\mathbf{\tilde{p}},\mathbf{\tilde{q}}}^{\mathbf{k},\mathbf{l},\mathbf{\tilde{k}},\mathbf{\tilde{l}}+\mathbf{e}_{\tilde{n}_{\mathrm{I}}}}.
\end{equation}

With the help of Eqs.~\ref{eq:eq13}-\ref{eq:eq16}, we finally obtain the HEOM as follows
\begin{equation}\label{eq:eq17}
\begin{split}
\partial_{t}\hat{\varrho}_{\mathbf{p},\mathbf{q},\mathbf{\tilde{p}},\mathbf{\tilde{q}}}^{\mathbf{k},\mathbf{l},\mathbf{\tilde{k}},\mathbf{\tilde{l}}}=&-i\Big{[}\hat{H}_{\mathrm{s}},\hat{\varrho}_{\mathbf{p},\mathbf{q},\mathbf{\tilde{p}},\mathbf{\tilde{q}}}^{\mathbf{k},\mathbf{l},\mathbf{\tilde{k}},\mathbf{\tilde{l}}}\Big{]}-(\mathbf{p}\cdot\mathbf{v_{R}}+\mathbf{k}\cdot \mathbf{v_{R}^{*}}+\mathbf{q}\cdot \mathbf{v_{I}}+\mathbf{l}\cdot \mathbf{v_{I}^{*}}+\mathbf{\tilde{p}}\cdot\mathbf{\tilde{v}_{R}}+\mathbf{\tilde{k}}\cdot\mathbf{\tilde{v}_{R}^{*}}+\mathbf{\tilde{q}}\cdot\mathbf{\tilde{v}_{I}}+\mathbf{\tilde{l}}\cdot\mathbf{\tilde{v}_{I}^{*}})\hat{\varrho}_{\mathbf{p},\mathbf{q},\mathbf{\tilde{p}},\mathbf{\tilde{q}}}^{\mathbf{k},\mathbf{l},\mathbf{\tilde{k}},\mathbf{\tilde{l}}}\\
+&\hat{S}\Big{(}\sum_{n_{\mathrm{R}}}P_{n_{\mathrm{R}}}U_{n_{\mathrm{R}}}^{\mathrm{R}}\hat{\varrho}_{\mathbf{p}-\mathbf{e}_{n_{\mathrm{R}}},\mathbf{q},\mathbf{\tilde{p}},\mathbf{\tilde{q}}}^{\mathbf{k},\mathbf{l},\mathbf{\tilde{k}},\mathbf{\tilde{l}}}+\sum_{n_{\mathrm{I}}}Q_{n_{\mathrm{I}}}U_{n_{\mathrm{I}}}^{\mathrm{I}}\hat{\varrho}_{\mathbf{p},\mathbf{q}-\mathbf{e}_{n_{\mathrm{I}}},\mathbf{\tilde{p}},\mathbf{\tilde{q}}}^{\mathbf{k},\mathbf{l},\mathbf{\tilde{k}},\mathbf{\tilde{l}}}\Big{)}+\hat{S}^{\dag}\Big{(}\sum_{\tilde{n}_{\mathrm{R}}}\tilde{P}_{\tilde{n}_{\mathrm{R}}}\tilde{U}_{\tilde{n}_{\mathrm{R}}}^{\mathrm{R}}\hat{\varrho}_{\mathbf{p},\mathbf{q},\mathbf{\tilde{p}}-\mathbf{e}_{\tilde{n}_{\mathrm{R}}},\mathbf{\tilde{q}}}^{\mathbf{k},\mathbf{l},\mathbf{\tilde{k}},\mathbf{\tilde{l}}}+\sum_{\tilde{n}_{\mathrm{I}}}\tilde{Q}_{\tilde{n}_{\mathrm{I}}}\tilde{U}_{\tilde{n}_{\mathrm{I}}}^{\mathrm{I}}\hat{\varrho}_{\mathbf{p},\mathbf{q},\mathbf{\tilde{p}},\mathbf{\tilde{q}}-\mathbf{e}_{\tilde{n}_{\mathrm{I}}}}^{\mathbf{k},\mathbf{l},\mathbf{\tilde{k}},\mathbf{\tilde{l}}}\Big{)}\\
+&\Big{(}\sum_{n_{\mathrm{R}}}K_{n_{\mathrm{R}}}U_{n_{\mathrm{R}}}^{\mathrm{R}*}\hat{\varrho}_{\mathbf{p},\mathbf{q},\mathbf{\tilde{p}},\mathbf{\tilde{q}}}^{\mathbf{k}-\mathbf{e}_{n_{\mathrm{R}}},\mathbf{l},\mathbf{\tilde{k}},\mathbf{\tilde{l}}}+\sum_{n_{\mathrm{I}}}L_{n_{\mathrm{I}}}U_{n_{\mathrm{I}}}^{\mathrm{I}*}\hat{\varrho}_{\mathbf{p},\mathbf{q},\mathbf{\tilde{p}},\mathbf{\tilde{q}}}^{\mathbf{k},\mathbf{l}-\mathbf{e}_{n_{\mathrm{I}}},\mathbf{\tilde{k}},\mathbf{\tilde{l}}}\Big{)}\hat{S}^{\dagger}+\Big{(}\sum_{\tilde{n}_{\mathrm{R}}}\tilde{K}_{\tilde{n}_{\mathrm{R}}}\tilde{U}_{\tilde{n}_{\mathrm{R}}}^{\mathrm{R}*}\hat{\varrho}_{\mathbf{p},\mathbf{q},\mathbf{\tilde{p}},\mathbf{\tilde{q}}}^{\mathbf{k},\mathbf{l},\mathbf{\tilde{k}}-\mathbf{e}_{\tilde{n}_{\mathrm{R}}},\mathbf{\tilde{l}}}+\sum_{\tilde{n}_{\mathrm{I}}}\tilde{L}_{\tilde{n}_{\mathrm{I}}}\tilde{U}_{\tilde{n}_{\mathrm{I}}}^{\mathrm{I}*}\hat{\varrho}_{\mathbf{p},\mathbf{q},\mathbf{\tilde{p}},\mathbf{\tilde{q}}}^{\mathbf{k},\mathbf{l},\mathbf{\tilde{k}},\mathbf{\tilde{l}}-\mathbf{e}_{\tilde{n}_{\mathrm{I}}}}\Big{)}\hat{S}\\
-&\Big{[}\hat{S}^{\dagger},\sum_{n_{\mathrm{R}}}\hat{\varrho}_{\mathbf{p}+\mathbf{e}_{n_{\mathrm{R}}},\mathbf{q},\mathbf{\tilde{p}},\mathbf{\tilde{q}}}^{\mathbf{k},\mathbf{l},\mathbf{\tilde{k}},\mathbf{\tilde{l}}}\Big{]}-\Big{[}\hat{S}^{\dagger},\sum_{n_{\mathrm{I}}}\hat{\varrho}_{\mathbf{p},\mathbf{q}+\mathbf{e}_{n_{\mathrm{I}}},\mathbf{\tilde{p}},\mathbf{\tilde{q}}}^{\mathbf{k},\mathbf{l},\mathbf{\tilde{k}},\mathbf{\tilde{l}}}\Big{]}-\Big{[}\hat{S},\sum_{\tilde{n}_{\mathrm{R}}}\hat{\varrho}_{\mathbf{p},\mathbf{q},\mathbf{\tilde{p}}+\mathbf{e}_{\tilde{n}_{\mathrm{R}}},\mathbf{\tilde{q}}}^{\mathbf{k},\mathbf{l},\mathbf{\tilde{k}},\mathbf{\tilde{l}}}\Big{]}-\Big{[}\hat{S},\sum_{\tilde{n}_{\mathrm{I}}}\hat{\varrho}_{\mathbf{p},\mathbf{q},\mathbf{\tilde{p}},\mathbf{\tilde{q}}+\mathbf{e}_{\tilde{n}_{\mathrm{I}}}}^{\mathbf{k},\mathbf{l},\mathbf{\tilde{k}},\mathbf{\tilde{l}}}\Big{]}\\
+&\Big{[}\hat{S},\sum_{n_{\mathrm{R}}}\hat{\varrho}_{\mathbf{p},\mathbf{q},\mathbf{\tilde{p}},\mathbf{\tilde{q}}}^{\mathbf{k}+\mathbf{e}_{n_{\mathrm{R}}},\mathbf{l},\mathbf{\tilde{k}},\mathbf{\tilde{l}}}\Big{]}+\Big{[}\hat{S},\sum_{n_{\mathrm{I}}}\hat{\varrho}_{\mathbf{p},\mathbf{q},\mathbf{\tilde{p}},\mathbf{\tilde{q}}}^{\mathbf{k},\mathbf{l}+\mathbf{e}_{n_{\mathrm{I}}},\mathbf{\tilde{k}},\mathbf{\tilde{l}}}\Big{]}+\Big{[}\hat{S}^{\dagger},\sum_{\tilde{n}_{\mathrm{R}}}\hat{\varrho}_{\mathbf{p},\mathbf{q},\mathbf{\tilde{p}},\mathbf{\tilde{q}}}^{\mathbf{k},\mathbf{l},\mathbf{\tilde{k}}+\mathbf{e}_{\tilde{n}_{\mathrm{R}}},\mathbf{\tilde{l}}}\Big{]}+\Big{[}\hat{S}^{\dagger},\sum_{\tilde{n}_{\mathrm{I}}}\hat{\varrho}_{\mathbf{p},\mathbf{q},\mathbf{\tilde{p}},\mathbf{\tilde{q}}}^{\mathbf{k},\mathbf{l},\mathbf{\tilde{k}},\mathbf{\tilde{l}}+\mathbf{e}_{\tilde{n}_{\mathrm{I}}}}\Big{]},
\end{split}
\end{equation}
where $[\hat{\phi},\hat{\varphi}]\equiv\hat{\phi}\hat{\varphi}-\hat{\varphi}\hat{\phi}$. And Eq.~\ref{eq:eq17} is the main result of our paper. In fact, by combining the indexes as follows: $\mathbf{\Lambda}\equiv \mathbf{v}_{\mathbf{R}}\oplus\mathbf{v}_{\mathbf{I}}=\{\Lambda_{n}\}_{1}^{N_{\mathrm{R}}+N_{\mathrm{I}}}$, $\mathbf{\tilde{\Lambda}}\equiv \mathbf{\tilde{v}}_{\mathbf{R}}\oplus\mathbf{\tilde{v}}_{\mathbf{I}}=\{\tilde{\Lambda}_{\tilde{n}}\}_{1}^{\tilde{N}_{\mathrm{R}}+\tilde{N}_{\mathrm{I}}}$, $\mathbf{\Xi}\equiv \mathbf{u}_{\mathbf{R}}\oplus\mathbf{u}_{\mathbf{I}}=\{\Xi_{n}\}_{1}^{N_{\mathrm{R}}+N_{\mathrm{I}}}$, $\mathbf{\tilde{\Xi}}\equiv \mathbf{\tilde{u}}_{\mathbf{R}}\oplus\mathbf{\tilde{u}}_{\mathbf{I}}=\{\tilde{\Xi}_{\tilde{n}}\}_{1}^{\tilde{N}_{\mathrm{R}}+\tilde{N}_{\mathrm{I}}}$, $\mathbf{i}\equiv\mathbf{k}\oplus\mathbf{l}=\{I_{n}\}_{1}^{N_{\mathrm{R}}+N_{\mathrm{I}}}$, $\mathbf{j}\equiv\mathbf{p}\oplus\mathbf{q}=\{J_{n}\}_{1}^{N_{\mathrm{R}}+N_{\mathrm{I}}}$, $\mathbf{\tilde{i}}\equiv\mathbf{\tilde{k}}\oplus\mathbf{\tilde{l}}=\{\tilde{I}_{\tilde{n}}\}_{1}^{\tilde{N}_{\mathrm{R}}+\tilde{N}_{\mathrm{I}}}$, and $\mathbf{\tilde{j}}\equiv\mathbf{\tilde{p}}\oplus\mathbf{\tilde{q}}=\{\tilde{J}_{\tilde{n}}\}_{1}^{\tilde{N}_{\mathrm{R}}+\tilde{N}_{\mathrm{I}}}$, one can greatly simplify the expression of Eq.~\ref{eq:eq17},
\begin{equation*}
\begin{split}
\partial_{t}\hat{\varrho}_{\mathbf{j},\mathbf{\tilde{j}}}^{\mathbf{i},\mathbf{\tilde{i}}}=&-i\Big{[}\hat{H}_{\mathrm{s}},\hat{\varrho}_{\mathbf{j},\mathbf{\tilde{j}}}^{\mathbf{i},\mathbf{\tilde{i}}}\Big{]}-(\mathbf{j}\cdot\mathbf{\Lambda}+\mathbf{i}\cdot \mathbf{\Lambda^{*}}+\mathbf{\tilde{j}}\cdot\mathbf{\tilde{\Lambda}}+\mathbf{\tilde{i}}\cdot\mathbf{\tilde{\Lambda}^{*}})\hat{\varrho}_{\mathbf{j},\mathbf{\tilde{j}}}^{\mathbf{i},\mathbf{\tilde{i}}}\\
&+\hat{S}\sum_{n}J_{n}\Xi_{n}\hat{\varrho}_{\mathbf{j}-\mathbf{e}_{n},\mathbf{\tilde{j}}}^{\mathbf{i},\mathbf{\tilde{i}}}+\hat{S}^{\dag}\sum_{\tilde{n}}\tilde{J}_{\tilde{n}}\tilde{\Xi}_{\tilde{n}}\hat{\varrho}_{\mathbf{j},\mathbf{\tilde{j}}-\mathbf{e}_{\tilde{n}}}^{\mathbf{i},\mathbf{\tilde{i}}}+\sum_{n}I_{n}\Xi_{n}^{*}\hat{\varrho}_{\mathbf{j},\mathbf{\tilde{j}}}^{\mathbf{i}-\mathbf{e}_{n},\mathbf{\tilde{i}}}\hat{S}^{\dagger}+\sum_{\tilde{n}}\tilde{I}_{\tilde{n}}\tilde{\Xi}_{\tilde{n}}^{*}\hat{\varrho}_{\mathbf{j},\mathbf{\tilde{j}}}^{\mathbf{i},\mathbf{\tilde{i}}-\mathbf{e}_{\tilde{n}}}\hat{S}\\
&-\Big{[}\hat{S}^{\dagger},\sum_{n}\hat{\varrho}_{{\mathbf{j}+\mathbf{e}_{n}},\mathbf{\tilde{j}}}^{\mathbf{i},\mathbf{\tilde{i}}}\Big{]}-\Big{[}\hat{S},\sum_{\tilde{n}}\hat{\varrho}_{\mathbf{j},\mathbf{\tilde{j}}+\mathbf{e}_{\tilde{n}}}^{\mathbf{i},\mathbf{\tilde{i}}}\Big{]}+\Big{[}\hat{S},\sum_{n}\hat{\varrho}_{\mathbf{j},\mathbf{\tilde{j}}}^{{\mathbf{i}+\mathbf{e}_{n}},\mathbf{\tilde{i}}}\Big{]}+\Big{[}\hat{S}^{\dagger},\sum_{\tilde{n}}\hat{\varrho}_{\mathbf{j},\mathbf{\tilde{j}}}^{\mathbf{i},\mathbf{\tilde{i}}+\mathbf{e}_{\tilde{n}}}\Big{]}.
\end{split}
\end{equation*}
This result already resembles that in Refs.~\cite{33,35}. The initial-state conditions of the auxiliary matrices are $\hat{\varrho}_{\mathbf{0},\mathbf{0},\mathbf{0},\mathbf{0}}^{\mathbf{0},\mathbf{0},\mathbf{0},\mathbf{0}}=|\Phi(0)\rangle\langle\Phi(0)|$ and all the others are equal to zero. The HEOM consists of an infinite hierarchical equations and needs to be truncated. In spite of various truncation schemes have been proposed~\cite{33,34}, we adopt the standard method, i.e., setting all the auxiliary matrices, whose configuration indexes are larger than the cut-off number, to be zero. This scheme is sufficient and reliable, owing to the non-perturbative nature of the HEOM formalism. In a numerical simulation, we keep on adding the number of the hierarchy equations until the final result converges. It is necessary to point out that in the derivation of the HEOM, we did not use the usual Markovian approximation, the rotating-wave approximation or the perturbative approximation. In this sense, HEOM can be regarded as a rigorous numerical method.

\subsection{SD approach}\label{sec:sec2b}

In Subsec.~\ref{sec:sec2a}, we rigorously constructed the HEOM of a generalized Hamiltonian given by Eq.~\ref{eq:eq1} in the framework of the NMQSD method. Our result indicates that a deterministic HEOM can be extracted by a stochastic dynamical description. In addition to the NMSQD method, the SD scheme is another stochastic dynamics approach and also provides a way to understand the stochastic formulation of quantum dissipative systems. In this subsection, we shall derive the HEOM from the SD perspective.

The dynamics of the whole system given by Eq.~\ref{eq:eq1} is governed by the following quantum Liouville equation
\begin{equation}\label{eq:eq18}
\frac{d}{dt}\hat{\varrho}_{\mathrm{sb}}(t)=-i[\hat{H},\hat{\varrho}_{\mathrm{sb}}(t)].
\end{equation}
Many techniques have been proposed to avoid a direct computation of Eq.~\ref{eq:eq18} because of its intractability. The complexity of dissipative dynamics lies in the coupling between the quantum subsystem and the bath. It would be desirable to decouple the interaction in such a way that the evolution of the bath will no longer be explicitly involved in the evolution of the quantum subsystem. Such a decoupling scheme can be achieved by making use of the Hubbard-Stratonovich transformation, or alternatively, the It$\mathrm{\hat{o}}$ calculus~\cite{23,24,25}. The sacrifice of this decoupling scheme is to introduce auxiliary stochastic noises in the evolution of the quantum subsystem. As a result, the density matrix of the whole system, namely $\hat{\varrho}_{\mathrm{sb}}(t)$, can be expressed as (see Refs.~\cite{23,24,25} or Appendix~\ref{sec:secappa} for more details)
\begin{equation}\label{eq:eq19}
\hat{\varrho}_{\mathrm{sb}}(t)=\mathcal{M}\{\hat{\rho}_{\mathrm{s}}(t)\hat{\rho}_{\mathrm{b}}(t)\},
\end{equation}
where we have assumed the whole system is initially prepared in a product state $\hat{\varrho}_{\mathrm{sb}}(0)=\hat{\rho}_{\mathrm{s}}(0)\otimes\hat{\rho}_{\mathrm{b}}(0)$ with $\hat{\rho}_{\mathrm{b}}(0)=\hat{\rho}_{\mathrm{th}}$. The notation $\mathcal{M}\{...\}$ is the ensemble mean over the stochastic noises. The stochastic density matrices $\hat{\rho}_{\mathrm{s}}(t)$ and $\hat{\rho}_{\mathrm{b}}(t)$ obey the following stochastic differential equations, respectively
\begin{equation}\label{eq:eq20}
\begin{split}
d\hat{\rho}_{\mathrm{s}}(t)=&-i[\hat{H}_{\mathrm{s}},\hat{\rho}_{\mathrm{s}}(t)]dt-\frac{i}{2}[\hat{S},\hat{\rho}_{\mathrm{s}}(t)]dw_{11}-\frac{i}{2}[\hat{S}^{\dag},\hat{\rho}_{\mathrm{s}}(t)]dw_{12}\\
&+\frac{1}{2}\{\hat{S},\hat{\rho}_{\mathrm{s}}(t)\}dw_{21}^{*}+\frac{1}{2}\{\hat{S}^{\dagger},\hat{\rho}_{\mathrm{s}}(t)\}dw_{22}^{*},
\end{split}
\end{equation}
\begin{equation}\label{eq:eq21}
\begin{split}
d\hat{\rho}_{\mathrm{b}}(t)=&-i[\hat{H}_{\mathrm{b}},\hat{\rho}_{\mathrm{b}}(t)]dt-\frac{i}{2}[\hat{B}^{\dag},\hat{\rho}_{\mathrm{b}}(t)]dw_{21}-\frac{i}{2}[\hat{B},\hat{\rho}_{\mathrm{b}}(t)]dw_{22}\\
&+\frac{1}{2}\{\hat{B}^{\dag},\hat{\rho}_{\mathrm{b}}(t)\}dw_{11}^{*}+\frac{1}{2}\{\hat{B},\hat{\rho}_{\mathrm{b}}(t)\}dw_{12}^{*},
\end{split}
\end{equation}
where $\{\hat{\phi},\hat{\varphi}\}\equiv\hat{\phi}\hat{\varphi}+\hat{\varphi}\hat{\phi}$, $w_{1j}=w_{1j}(t)\equiv\int_{0}^{t}d\tau[\nu_{1j}(\tau)+i\nu_{4j}(\tau)]$ and $w_{2j}=w_{2j}(t)\equiv\int_{0}^{t}d\tau[\nu_{2j}(\tau)+i\nu_{3j}(\tau)]$ (with $j=1,2$) are four independent complex Wiener processes, and $\nu_{ij}(t)$ (with $i=1,2,3,4$) are independent Gaussian white noises which satisfy $\mathcal{M}\{\nu_{ij}(t)\}=0$ and $\mathcal{M}\{\nu_{ij}(t)\nu_{i'j'}(\tau)\}=\delta_{ii'}\delta_{jj'}\delta(t-\tau)$. Please note that, in this present subsection, all the stochastic differential and integral equations are in the It$\mathrm{\hat{o}}$ sense.

However, it should be stressed that we want to calculate the (stochastic) reduced density matrix which is defined by $\hat{\tilde{\rho}}_{\mathrm{s}}(t)\equiv\mathrm{tr}_{\mathrm{b}}[\hat{\rho}_{\mathrm{s}}(t)\hat{\rho}_{\mathrm{b}}(t)]= \hat{\rho}_{\mathrm{s}}(t)\mathrm{tr}_{\mathrm{b}}[\hat{\rho}_{\mathrm{b}}(t)]$. Employing a Girsanov transformation~\cite{23,24,25}, one can absorb $\mathrm{tr}_{\mathrm{b}}[\hat{\rho}_{\mathrm{b}}(t)]$ into the measures of the Wiener processes and obtain the stochastic Liouville
equation of $\hat{\tilde{\rho}}_{\mathrm{s}}(t)$ as follows~\cite{51}:
\begin{equation}\label{eq:eq22}
\begin{split}
d\hat{\tilde{\rho}}_{\mathrm{s}}(t)=&-i[\hat{H}_{\mathrm{s}}+\bar{g}_{1}(t)\hat{S}+\bar{g}_{2}(t)\hat{S}^{\dagger},\hat{\tilde{\rho}}_{\mathrm{s}}(t)]dt\\
&-\frac{i}{2}[\hat{S},\hat{\tilde{\rho}}_{\mathrm{s}}(t)]dw_{11}-\frac{i}{2}[\hat{S}^{\dag},\hat{\tilde{\rho}}_{\mathrm{s}}(t)]dw_{12}+\frac{1}{2}\{\hat{S},\hat{\tilde{\rho}}_{\mathrm{s}}(t)\}dw_{21}^{*}+\frac{1}{2}\{\hat{S}^{\dagger},\hat{\tilde{\rho}}_{\mathrm{s}}(t)\}dw_{22}^{*},
\end{split}
\end{equation}
here $\bar{g}_{1}(t)\equiv \mathrm{tr}_{\mathrm{b}}[\hat{B}^{\dag}\hat{\rho}_{\mathrm{b}}(t)]/\mathrm{tr}_{\mathrm{b}}[\hat{\rho}_{\mathrm{b}}(t)]$ and $\bar{g}_{2}(t)\equiv \mathrm{tr}_{\mathrm{b}}[\hat{B}\hat{\rho}_{\mathrm{b}}(t)]/\mathrm{tr}_{\mathrm{b}}[\hat{\rho}_{b}(t)]$ are the bath-induced mean fields which fully characterize the influences of the bath on the quantum subsystem, in fact, they play a similar role to that of the influence functional in the path-integral treatment~\cite{23}. The explicit expressions of $\bar{g}_{1,2}(t)$ can be determined by the stochastic evolution equation of the bath, i.e., Eq.~\ref{eq:eq21}, and the results are given by~\cite{51}
\begin{equation}\label{eq:eq23}
\bar{g}_{1}(t)=\frac{i}{2}\int_{0}^{t}d\tau\{\acute{\alpha}(t-\tau)[\nu_{22}(\tau)+i\nu_{32}(\tau)]-\grave{\alpha}(t-\tau)[i\nu_{12}(\tau)+\nu_{42}(\tau)]\},
\end{equation}
\begin{equation}\label{eq:eq24}
\bar{g}_{2}(t)=-\frac{i}{2}\int_{0}^{t}d\tau\{\acute{\alpha}^{*}(t-\tau)[\nu_{21}(\tau)+i\nu_{31}(\tau)]+\grave{\alpha}^{*}(t-\tau)[i\nu_{11}(\tau)+\nu_{41}(\tau)]\},
\end{equation}
where
\begin{equation}\label{eq:eq25}
\acute{\alpha}(t)\equiv\int_{0}^{\infty}d\omega J(\omega)e^{i\omega t},
\end{equation}
\begin{equation}\label{eq:eq26}
\grave{\alpha}(t)\equiv\int_{0}^{\infty}d\omega J(\omega)\coth\Bigg{(}\frac{\beta\omega}{2}\Bigg{)}e^{i\omega t}.
\end{equation}
By taking the statistical ensemble mean over all the Wiener processes, one can finally obtain the quantum master equation of the (deterministic) reduced density matrix $\hat{\varrho}_{\mathrm{s}}(t)\equiv\mathcal{M}\{\hat{\tilde{\rho}}_{\mathrm{s}}(t)\}$ as follows~\cite{51}
\begin{equation}\label{eq:eq27}
\begin{split}
\frac{d}{dt}\hat{\varrho}_{\mathrm{s}}(t)=&-i[\hat{H}_{\mathrm{s}},\hat{\varrho}_{\mathrm{s}}(t)]+\Bigg{[}\hat{S},\int_{0}^{t}d\tau\acute{\alpha}(t-\tau)\mathcal{M}\Big{\{}\big{[}\nu_{22}(\tau)\hat{\tilde{\rho}}_{s}(t)+i\nu_{32}(\tau)\hat{\tilde{\rho}}_{s}(t)\big{]}\Big{\}}\Bigg{]}\\
&-\Bigg{[}\hat{S},\int_{0}^{t}d\tau\grave{\alpha}(t-\tau)\mathcal{M}\Big{\{}\big{[}i\nu_{12}(\tau)\hat{\tilde{\rho}}_{s}(t)+\nu_{42}(\tau)\hat{\tilde{\rho}}_{s}(t)\big{]}\Big{\}}\Bigg{]}\\
&-\Bigg{[}\hat{S}^{\dagger},\int_{0}^{t}d\tau\acute{\alpha}^{*}(t-\tau)\mathcal{M}\Big{\{}\big{[}\nu_{21}(\tau)\hat{\tilde{\rho}}_{s}(t)+i\nu_{31}(\tau)\hat{\tilde{\rho}}_{s}(t)\big{]}\Big{\}}\Bigg{]}\\
&-\Bigg{[}\hat{S}^{\dag},\int_{0}^{t}d\tau\grave{\alpha}^{*}(t-\tau)\mathcal{M}\Big{\{}\big{[}i\nu_{11}(\tau)\hat{\tilde{\rho}}_{s}(t)+\nu_{41}(\tau)\hat{\tilde{\rho}}_{s}(t)\big{]}\Big{\}}\Bigg{]}.
\end{split}
\end{equation}
Here, in the derivation of Eq.~\ref{eq:eq27}, we have used the non-anticipating property which states that $\hat{\tilde{\rho}}_{s}(t)$ is independent of arbitrary Wiener increments, i.e.,
\begin{equation*}
\mathcal{M}\{\hat{\tilde{\rho}}_{s}(t)dw_{jj'}(t)\}=0.
\end{equation*}

It is easy to check that $\acute{\alpha}(t)=\alpha^{*}(t)-\tilde{\alpha}(t)$ and $\grave{\alpha}(t)=\alpha^{*}(t)+\tilde{\alpha}(t)$, and one can reexpress $\bar{g}_{1,2}(t)$ in term of $\alpha(t)$ and $\tilde{\alpha}(t)$. Then, one can rewritten Eq.~\ref{eq:eq27} as
\begin{equation}\label{eq:eq28}
\begin{split}
\frac{d}{dt}\hat{\varrho}_{\mathrm{s}}(t)=&-i[\hat{H}_{\mathrm{s}},\hat{\varrho}_{\mathrm{s}}(t)]-\Bigg{[}\hat{S}^{\dagger},\mathcal{M}\Big{\{}\Big{(}\sum_{n_{\mathrm{R}}}\mathcal{\hat{P}}_{n_{\mathrm{R}}}^{\mathrm{R}}+\sum_{n_{\mathrm{I}}}\mathcal{\hat{P}}_{n_{\mathrm{I}}}^{\mathrm{I}}\Big{)}\hat{\tilde{\rho}}_{\mathrm{s}}(t)\Big{\}}\Bigg{]}-\Bigg{[}\hat{S},\mathcal{M}\Big{\{}\Big{(}\sum_{\tilde{n}_{\mathrm{R}}}\mathcal{\hat{Q}}_{\tilde{n}_{\mathrm{R}}}^{\mathrm{R}}+\sum_{\tilde{n}_{\mathrm{I}}}\mathcal{\hat{Q}}_{\tilde{n}_{\mathrm{I}}}^{\mathrm{I}}\Big{)}\hat{\tilde{\rho}}_{\mathrm{s}}(t)\Big{\}}\Bigg{]}\\
&+\Bigg{[}\hat{S},\mathcal{M}\Big{\{}\Big{(}\sum_{n_{\mathrm{R}}}\mathcal{\hat{K}}_{n_{\mathrm{R}}}^{\mathrm{R}}+\sum_{n_{\mathrm{I}}}\mathcal{\hat{K}}_{n_{\mathrm{I}}}^{\mathrm{I}}\Big{)}\hat{\tilde{\rho}}_{\mathrm{s}}(t)\Big{\}}\Bigg{]}+\Bigg{[}\hat{S}^{\dagger},\mathcal{M}\Big{\{}\Big{(}\sum_{\tilde{n}_{\mathrm{R}}}\mathcal{\hat{L}}_{\tilde{n}_{\mathrm{R}}}^{\mathrm{R}}+\sum_{\tilde{n}_{\mathrm{I}}}\mathcal{\hat{L}}_{\tilde{n}_{\mathrm{I}}}^{\mathrm{I}}\Big{)}\hat{\tilde{\rho}}_{\mathrm{s}}(t)\Big{\}}\Bigg{]},
\end{split}
\end{equation}
where the super-operators are defined by
\begin{equation}\label{eq:eq29}
\mathcal{\hat{P}}_{n_{\mathrm{X}}}^{\mathrm{X}}\equiv\frac{1}{2}(i)^{\theta_{\mathrm{X}}}\int_{0}^{t}d\tau U_{n_{\mathrm{X}}}^{\mathrm{X}}\exp[-V_{n_{\mathrm{X}}}^{\mathrm{X}}(t-\tau)][\nu_{21}(\tau)+i\nu_{31}(\tau)+i\nu_{11}(\tau)+\nu_{41}(\tau)],
\end{equation}
\begin{equation}\label{eq:eq30}
\mathcal{\hat{Q}}_{\tilde{n}_{\mathrm{X}}}^{\mathrm{X}}=\frac{1}{2}(i)^{\theta_{\mathrm{X}}}\int_{0}^{t}d\tau \tilde{U}_{\tilde{n}_{\mathrm{X}}}^{\mathrm{X}}\exp[-\tilde{V}_{\tilde{n}_{\mathrm{X}}}^{\mathrm{X}}(t-\tau)][\nu_{22}(\tau)+i\nu_{32}(\tau)+i\nu_{12}(\tau)+\nu_{42}(\tau)],
\end{equation}
\begin{equation}\label{eq:eq31}
\mathcal{\hat{K}}_{n_{\mathrm{X}}}^{\mathrm{X}}=\frac{1}{2}(-i)^{\theta_{\mathrm{X}}}\int_{0}^{t}d\tau U_{n_{\mathrm{X}}}^{\mathrm{X}*}\exp[-V_{n_{\mathrm{X}}}^{\mathrm{X}*}(t-\tau)][\nu_{22}(\tau)+i\nu_{32}(\tau)-i\nu_{12}(\tau)-\nu_{42}(\tau)],
\end{equation}
\begin{equation}\label{eq:eq32}
\mathcal{\hat{L}}_{\tilde{n}_{\mathrm{X}}}^{\mathrm{X}}=\frac{1}{2}(-i)^{\theta_{\mathrm{X}}}\int_{0}^{t}d\tau \tilde{U}_{\tilde{n}_{\mathrm{X}}}^{\mathrm{X}*}\exp[-\tilde{V}_{\tilde{n}_{\mathrm{X}}}^{\mathrm{X}*}(t-\tau)][\nu_{21}(\tau)+i\nu_{31}(\tau)-i\nu_{11}(\tau)-\nu_{41}(\tau)].
\end{equation}

Similar to the procedure displayed in Sec.~\ref{sec:sec2a}, we would like to rewritten Eq.~\ref{eq:eq28} as
\begin{equation*}
\begin{split}
\partial_{t}\hat{\varrho}_{\mathbf{0},\mathbf{0},\mathbf{0},\mathbf{0}}^{\mathbf{0},\mathbf{0},\mathbf{0},\mathbf{0}}=&-i\Big{[}\hat{H}_{\mathrm{s}},\hat{\varrho}_{\mathbf{0},\mathbf{0},\mathbf{0},\mathbf{0}}^{\mathbf{0},\mathbf{0},\mathbf{0},\mathbf{0}}\Big{]}\\
&-\Big{[}\hat{S}^{\dag},\sum_{n_{\mathrm{R}}}\hat{\varrho}_{0...1_{n_{\mathrm{R}}}...0,\mathbf{0},\mathbf{0},\mathbf{0}}^{\mathbf{0},\mathbf{0},\mathbf{0},\mathbf{0}}+\sum_{n_{\mathrm{I}}}\hat{\varrho}_{\mathbf{0},0...1_{n_{\mathrm{I}}}...0,\mathbf{0},\mathbf{0}}^{\mathbf{0},\mathbf{0},\mathbf{0},\mathbf{0}}\Big{]}-\Big{[}\hat{S},\sum_{\tilde{n}_{\mathrm{R}}}\hat{\varrho}_{\mathbf{0},\mathbf{0},0...1_{\tilde{n}_{\mathrm{R}}}...0,\mathbf{0}}^{\mathbf{0},\mathbf{0},\mathbf{0},\mathbf{0}}+\sum_{\tilde{n}_{\mathrm{I}}}\hat{\varrho}_{\mathbf{0},\mathbf{0},\mathbf{0},0...1_{\tilde{n}_{\mathrm{I}}}...0}^{\mathbf{0},\mathbf{0},\mathbf{0},\mathbf{0}}\Big{]}\\
&+\Big{[}\hat{S},\sum_{n_{\mathrm{R}}}\hat{\varrho}_{\mathbf{0},\mathbf{0},\mathbf{0},\mathbf{0}}^{0...1_{n_{\mathrm{R}}}...0,\mathbf{0},\mathbf{0},\mathbf{0}}+\sum_{n_{\mathrm{I}}}\hat{\varrho}_{\mathbf{0},\mathbf{0},\mathbf{0},\mathbf{0}}^{\mathbf{0},0...1_{n_{\mathrm{I}}}...0,\mathbf{0},\mathbf{0}}\Big{]}+\Big{[}\hat{S}^{\dagger},\sum_{\tilde{n}_{\mathrm{R}}}\hat{\varrho}_{\mathbf{0},\mathbf{0},\mathbf{0},\mathbf{0}}^{\mathbf{0},\mathbf{0},0...1_{\tilde{n}_{\mathrm{R}}}...0,\mathbf{0}}+\sum_{\tilde{n}_{\mathrm{I}}}\hat{\varrho}_{\mathbf{0},\mathbf{0},\mathbf{0},\mathbf{0}}^{\mathbf{0},\mathbf{0},\mathbf{0},0...1_{\tilde{n}_{\mathrm{I}}}...0}\Big{]},
\end{split}
\end{equation*}
where we have defined the expressions of the auxiliary matrices in the SD scheme as follows
\begin{equation}\label{eq:eq33}
\begin{split}
\hat{\varrho}_{\mathbf{p},\mathbf{q},\mathbf{\tilde{p}},\mathbf{\tilde{q}}}^{\mathbf{k},\mathbf{l},\mathbf{\tilde{k}},\mathbf{\tilde{l}}}=&\hat{\varrho}_{\mathbf{p},\mathbf{q},\mathbf{\tilde{p}},\mathbf{\tilde{q}}}^{\mathbf{k},\mathbf{l},\mathbf{\tilde{k}},\mathbf{\tilde{l}}}(t)\equiv\mathcal{M}\big{\{}\hat{\tilde{\rho}}_{\mathbf{p},\mathbf{q},\mathbf{\tilde{p}},\mathbf{\tilde{q}}}^{\mathbf{k},\mathbf{l},\mathbf{\tilde{k}},\mathbf{\tilde{l}}}(t)\big{\}}=\mathcal{M}\big{\{}\hat{\tilde{\rho}}_{\mathbf{p},\mathbf{q},\mathbf{\tilde{p}},\mathbf{\tilde{q}}}^{\mathbf{k},\mathbf{l},\mathbf{\tilde{k}},\mathbf{\tilde{l}}}\big{\}}\\
=&\mathcal{M}\Big{\{}(\mathcal{\hat{P}}_{n_{\mathrm{R}}}^{\mathrm{R}})^{\mathrm{p}_{n_{\mathrm{R}}}}(\mathcal{\hat{P}}_{n_{\mathrm{I}}}^{\mathrm{I}})^{\mathrm{q}_{n_{\mathrm{I}}}}(\mathcal{\hat{Q}}_{\tilde{n}_{\mathrm{R}}}^{\mathrm{R}})^{\tilde{\mathrm{p}}_{\tilde{n}_{\mathrm{R}}}}(\mathcal{\hat{Q}}_{\tilde{n}_{\mathrm{I}}}^{\mathrm{I}})^{\tilde{\mathrm{q}}_{\tilde{n}_{\mathrm{I}}}}(\mathcal{\hat{K}}_{n_{\mathrm{R}}}^{\mathrm{R}})^{\mathrm{k}_{n_{\mathrm{R}}}}(\mathcal{\hat{K}}_{n_{\mathrm{I}}}^{\mathrm{I}})^{\mathrm{l}_{n_{\mathrm{I}}}}(\mathcal{\hat{L}}_{\tilde{n}_{\mathrm{R}}}^{\mathrm{R}})^{\tilde{\mathrm{k}}_{\tilde{n}_{\mathrm{R}}}}(\mathcal{\hat{L}}_{\tilde{n}_{\mathrm{I}}}^{\mathrm{I}})^{\tilde{\mathrm{l}}_{\tilde{n}_{\mathrm{I}}}}\hat{\tilde{\rho}}_{s}(t)\Big{\}}.
\end{split}
\end{equation}
Taking the time derivative of the auxiliary matrix $\hat{\varrho}_{0...1_{n_{\mathrm{R}}}...0,\mathbf{0},\mathbf{0},\mathbf{0}}^{\mathbf{0},\mathbf{0},\mathbf{0},\mathbf{0}}$, one can find
\begin{equation*}
\begin{split}
\partial_{t}\hat{\varrho}_{0...1_{n_{\mathrm{R}}}...0,\mathbf{0},\mathbf{0},\mathbf{0}}^{\mathbf{0},\mathbf{0},\mathbf{0},\mathbf{0}}=&\mathcal{M}\Big{\{}\partial_{t}\big{(}\mathcal{\hat{P}}_{n_{\mathrm{R}}}^{\mathrm{R}}\hat{\tilde{\rho}}_{\mathbf{0},\mathbf{0},\mathbf{0},\mathbf{0}}^{\mathbf{0},\mathbf{0},\mathbf{0},\mathbf{0}}\big{)}\Big{\}}\\
=&\mathcal{M}\Big{\{}\big{(}\partial_{t}\mathcal{\hat{P}}_{n_{\mathrm{R}}}^{\mathrm{R}}\big{)}\hat{\tilde{\rho}}_{\mathbf{0},\mathbf{0},\mathbf{0},\mathbf{0}}^{\mathbf{0},\mathbf{0},\mathbf{0},\mathbf{0}}\Big{\}}+\mathcal{M}\Big{\{}\mathcal{\hat{P}}_{n_{\mathrm{R}}}^{\mathrm{R}}\big{(}\partial_{t}\hat{\tilde{\rho}}_{\mathbf{0},\mathbf{0},\mathbf{0},\mathbf{0}}^{\mathbf{0},\mathbf{0},\mathbf{0},\mathbf{0}}\big{)}\Big{\}}\\
=&-V_{n_{\mathrm{R}}}^{\mathrm{R}}\mathcal{M}\Big{\{}\mathcal{\hat{P}}_{n_{\mathrm{R}}}^{\mathrm{R}}\hat{\tilde{\rho}}_{\mathbf{0},\mathbf{0},\mathbf{0},\mathbf{0}}^{\mathbf{0},\mathbf{0},\mathbf{0},\mathbf{0}}\Big{\}}+\frac{1}{2}U_{n_{\mathrm{R}}}^{\mathrm{R}}\mathcal{M}\Big{\{}\big{[}\nu_{21}(t)+i\nu_{31}(t)+i\nu_{11}(t)+\nu_{41}(t)\big{]}\hat{\tilde{\rho}}_{\mathbf{0},\mathbf{0},\mathbf{0},\mathbf{0}}^{\mathbf{0},\mathbf{0},\mathbf{0},\mathbf{0}}\Big{\}}\\
&-i\Big{[}\hat{H}_{\mathrm{s}},\mathcal{M}\big{\{}\mathcal{\hat{P}}_{n_{\mathrm{R}}}^{\mathrm{R}}\hat{\tilde{\rho}}_{\mathbf{0},\mathbf{0},\mathbf{0},\mathbf{0}}^{\mathbf{0},\mathbf{0},\mathbf{0},\mathbf{0}}\big{\}}\Big{]}-\Big{[}\hat{S}^{\dag},\sum_{n_{\mathrm{R}}'}\mathcal{M}\big{\{}\mathcal{\hat{P}}_{n_{\mathrm{R}}}^{\mathrm{R}}\hat{\tilde{\rho}}_{0...1_{n_{\mathrm{R}}'}...0,\mathbf{0},\mathbf{0},\mathbf{0}}^{\mathbf{0},\mathbf{0},\mathbf{0},\mathbf{0}}\big{\}}+\sum_{n_{\mathrm{I}}}\mathcal{M}\big{\{}\mathcal{\hat{P}}_{n_{\mathrm{R}}}^{\mathrm{R}}\hat{\tilde{\rho}}_{\mathbf{0},0...1_{n_{\mathrm{I}}}...0,\mathbf{0},\mathbf{0}}^{\mathbf{0},\mathbf{0},\mathbf{0},\mathbf{0}}\big{\}}\Big{]}\\
&-\Big{[}\hat{S},\sum_{\tilde{n}_{R}}\mathcal{M}\big{\{}\mathcal{\hat{P}}_{n_{\mathrm{R}}}^{\mathrm{R}}\hat{\tilde{\rho}}_{\mathbf{0},\mathbf{0},0...1_{\tilde{n}_{\mathrm{R}}}...0,\mathbf{0}}^{\mathbf{0},\mathbf{0},\mathbf{0},\mathbf{0}}\big{\}}+\sum_{\tilde{n}_{\mathrm{I}}}\mathcal{M}\big{\{}\mathcal{\hat{P}}_{\tilde{n}_{\mathrm{R}}}^{\mathrm{R}}\hat{\tilde{\rho}}_{\mathbf{0},\mathbf{0},\mathbf{0},0...1_{\tilde{n}}...0}^{\mathbf{0},\mathbf{0},\mathbf{0},\mathbf{0}}\big{\}}\Big{]}\\
&+\Big{[}\hat{S},\sum_{n_{\mathrm{R}}'}\mathcal{M}\big{\{}\mathcal{\hat{P}}_{n_{\mathrm{R}}}^{\mathrm{R}}\hat{\tilde{\rho}}_{\mathbf{0},\mathbf{0},\mathbf{0},\mathbf{0}}^{0...1_{n_{\mathrm{R}}'}...0,\mathbf{0},\mathbf{0},\mathbf{0}}\big{\}}+\sum_{n_{\mathrm{I}}}\mathcal{M}\big{\{}\mathcal{\hat{P}}_{n_{\mathrm{R}}}^{\mathrm{R}}\hat{\tilde{\rho}}_{\mathbf{0},\mathbf{0},\mathbf{0},\mathbf{0}}^{\mathbf{0},0...1_{n_{\mathrm{I}}}...0,\mathbf{0},\mathbf{0}}\big{\}}\Big{]}\\
&+\Big{[}\hat{S}^{\dagger},\sum_{\tilde{n}_{\mathrm{R}}}\mathcal{M}\big{\{}\mathcal{\hat{P}}_{n_{\mathrm{R}}}^{\mathrm{R}}\hat{\tilde{\rho}}_{\mathbf{0},\mathbf{0},\mathbf{0},\mathbf{0}}^{\mathbf{0},\mathbf{0},0...1_{\tilde{n}_{\mathrm{R}}}...0,\mathbf{0}}\big{\}}+\sum_{\tilde{n}_{\mathrm{I}}}\mathcal{M}\big{\{}\mathcal{\hat{P}}_{n_{\mathrm{R}}}^{\mathrm{R}}\hat{\tilde{\rho}}_{\mathbf{0},\mathbf{0},\mathbf{0},\mathbf{0}}^{\mathbf{0},\mathbf{0},\mathbf{0},0...1_{\tilde{n}_{\mathrm{I}}}...0}\big{\}}\Big{]}.
\end{split}
\end{equation*}
The above equation still contains noise terms which can be removed by making use of the Furutsu-Novikov theorem~\cite{52}
\begin{equation*}
\begin{split}
\mathcal{M}\big{\{}[\nu_{21}(t)+i\nu_{31}(t)+i\nu_{11}(t)+\nu_{41}(t)]\hat{\tilde{\rho}}_{\mathbf{0},\mathbf{0},\mathbf{0},\mathbf{0}}^{\mathbf{0},\mathbf{0},\mathbf{0},\mathbf{0}}\big{\}}=\mathcal{M}\Bigg{\{}\frac{\delta\hat{\tilde{\rho}}_{s}(t)}{\delta\nu_{21}(t)}+i\frac{\delta\hat{\tilde{\rho}}_{s}(t)}{\delta\nu_{31}(t)}+i\frac{\delta\hat{\tilde{\rho}}_{s}(t)}{\delta\nu_{11}(t)}+\frac{\delta\hat{\tilde{\rho}}_{s}(t)}{\delta\nu_{41}(t)}\Bigg{\}},
\end{split}
\end{equation*}
these functional differentials can be easily evaluated from Eq.~\ref{eq:eq20}, then the time evolution of $\hat{\varrho}_{0...1_{n_{\mathrm{R}}}...0,\mathbf{0},\mathbf{0},\mathbf{0}}^{\mathbf{0},\mathbf{0},\mathbf{0},\mathbf{0}}$ can be obtained as follows
\begin{equation}\label{eq:eq34}
\begin{split}
\partial_{t}\hat{\varrho}_{0...1_{n_{\mathrm{R}}}...0,\mathbf{0},\mathbf{0},\mathbf{0}}^{\mathbf{0},\mathbf{0},\mathbf{0},\mathbf{0}}=&-i\Big{[}\hat{H}_{\mathrm{s}},\hat{\varrho}_{0...1_{n_{\mathrm{R}}}...0,\mathbf{0},\mathbf{0},\mathbf{0}}^{\mathbf{0},\mathbf{0},\mathbf{0},\mathbf{0}}\Big{]}-V_{n_{\mathrm{R}}}^{\mathrm{R}}\hat{\varrho}_{0...1_{n_{\mathrm{R}}}...0,\mathbf{0},\mathbf{0},\mathbf{0}}^{\mathbf{0},\mathbf{0},\mathbf{0},\mathbf{0}}+U_{n_{\mathrm{R}}}^{\mathrm{R}}\hat{S}\hat{\varrho}_{\mathbf{0},\mathbf{0},\mathbf{0},\mathbf{0}}^{\mathbf{0},\mathbf{0},\mathbf{0},\mathbf{0}}\\
&-\Big{[}\hat{S}^{\dag},\sum_{n_{\mathrm{R}}}\hat{\varrho}_{0...1_{n_{\mathrm{R}}}...1_{n_{\mathrm{R}}'}...0,\mathbf{0},\mathbf{0},\mathbf{0}}^{\mathbf{0},\mathbf{0},\mathbf{0},\mathbf{0}}+\sum_{n_{\mathrm{I}}}\hat{\varrho}_{0...1_{n_{\mathrm{R}}}...0,0...1_{n_{\mathrm{I}}}...0,\mathbf{0},\mathbf{0}}^{\mathbf{0},\mathbf{0},\mathbf{0},\mathbf{0}}\Big{]}\\
&-\Big{[}\hat{S},\sum_{\tilde{n}_{\mathrm{R}}}\hat{\varrho}_{0...1_{n_{\mathrm{R}}}...0,\mathbf{0},0...1_{\tilde{n}_{\mathrm{R}}}...0,\mathbf{0}}^{\mathbf{0},\mathbf{0},\mathbf{0},\mathbf{0}}+\sum_{\tilde{n}_{\mathrm{I}}}\hat{\varrho}_{0...1_{n_{\mathrm{R}}}...0,\mathbf{0},\mathbf{0},0...1_{\tilde{n}_{\mathrm{I}}}...0}^{\mathbf{0},\mathbf{0},\mathbf{0},\mathbf{0}}\Big{]}\\
&+\Big{[}\hat{S},\sum_{n_{\mathrm{R}}}\hat{\varrho}_{0...1_{n_{\mathrm{R}}'}...0,\mathbf{0},\mathbf{0},\mathbf{0}}^{0...1_{n_{\mathrm{R}}}...0,\mathbf{0},\mathbf{0},\mathbf{0}}+\sum_{n_{\mathrm{I}}}\hat{\varrho}_{0...1_{n_{\mathrm{R}}}...0,\mathbf{0},\mathbf{0},\mathbf{0}}^{\mathbf{0},0...1_{n_{\mathrm{I}}}...0,\mathbf{0},\mathbf{0}}\Big{]}\\
&+\Big{[}\hat{S}^{\dagger},\sum_{\tilde{n}_{\mathrm{R}}}\hat{\varrho}_{0...1_{n_{\mathrm{R}}}...0,\mathbf{0},\mathbf{0},\mathbf{0}}^{\mathbf{0},\mathbf{0},0...1_{\tilde{n}_{\mathrm{R}}}...0,\mathbf{0}}+\sum_{\tilde{n}_{\mathrm{I}}}\hat{\varrho}_{0...1_{n_{\mathrm{R}}}...0,\mathbf{0},\mathbf{0},\mathbf{0}}^{\mathbf{0},\mathbf{0},\mathbf{0},0...1_{\tilde{n}_{\mathrm{I}}}...0}\Big{]}.
\end{split}
\end{equation}
By repeating the same procedure for other auxiliary matrices, one can obtain the same HEOM with Eq.~\ref{eq:eq17}.

Before moving on to the next section, we would like to make some remarks on the NMQSD and the SD schemes. As discussed above, although we have proved that these two frameworks give the identical HEOM, the NMQSD and SD methods are completely different. The stochastic Schr$\mathrm{\ddot{o}}$dinger equation in Eq.~\ref{eq:eq2} derived by the NMQSD method still contains two undetermined functional-derivative terms, in practice, it is impossible to perform straightforwardly the stochastic simulation of Eq.~\ref{eq:eq2} except for a very few examples where the operators $\mathcal{\hat{O}}(t,\tau,\mathbf{z}_{t}^{*},\mathbf{\tilde{z}}_{t}^{*})$ and $\mathcal{\hat{G}}(t,\tau,\mathbf{z}_{t}^{*},\mathbf{\tilde{z}}_{t}^{*})$ can be (or at least approximately) determined. On the contrary, the stochastic Liouville equation in Eq.~\ref{eq:eq22} is a closed equation of motion, one can generate the bath-induced random fields $\bar{g}_{1,2}(t)$ by the convolution method~\cite{53} and perform directly the the stochastic simulations of Eq.~\ref{eq:eq22}. In fact, one can demonstrate that the stochastic Liouville equation given by Eq.~\ref{eq:eq22} can be recast into the stochastic differential equation suggested by Stockburger and Grabert~\cite{54} by making a simple combination of the Wiener processes $w_{jj'}(t)$ and the bath-induced stochastic field $\bar{g}_{1,2}(t)$~\cite{27}. However, the stochastic simulation is not reliable in the long-time regime and is deemed to be less efficient than solving the deterministic quantum master equation because of the random fluctuations. Therefore, a deterministic approach is always preferred. That is the main reason why we choose the HEOM formulation, rather than stochastic simulations.

\section{Results}\label{sec:sec3}

In Sec.~\ref{sec:sec2}, we extracted the same HEOM from two different stochastic dynamical formulations, here, in this section, we would like to consider two special cases: the self-adjoint coupling operator case and the zero-temperature bath case. We compare our numerical results with some well-known results. Three important quantum dissipative models are considered as the illustrative examples in this section, namely the pure dephasing model in a finite-temperature bath~\cite{55,56}, the spontaneous decay of a two-level atom in a vacuum~\cite{57} and the famous spin-boson model without the rotating-wave approximation~\cite{14,16,57}. The dephasing model and the spontaneous decay model can be exactly solved and have many useful applications in quantum information and quantum optics fields~\cite{55,56,57}. There is no rigorous solution for the spin-boson model beyond the rotating-wave approximation. In this case, we compare our numerical results with the perturbative results obtained by the generalized Silbey-Harris transformation~\cite{14,16,radd1,radd2}.

%%%%%%%%%%%%%%%%%%%%%%%%
\begin{figure}
\centering
\includegraphics[angle=0,width=8cm]{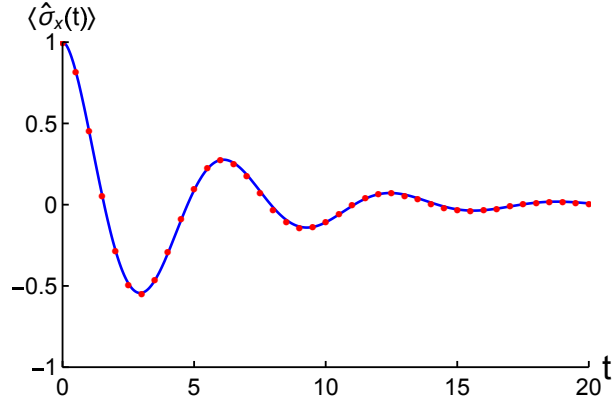}
\caption{\label{fig:fig1} The reduced dynamics of the pure dephasing model with initial state $|\Phi(0)\rangle=\frac{1}{\sqrt{2}}(|e\rangle+|g\rangle)$, where $|g\rangle$ and $|e\rangle$ denote the ground and the excited states of $\hat{\sigma}_{z}$, respectively. The blue solid line is the exactly analytical result and the red circles are the numerical results obtained by HEOM. Other parameters are chosen as $\chi=0.002$, $\omega_{c}=5$, $\beta=0.015$ and $\omega_{0}=1$.}
\end{figure}
%%%%%%%%%%%%%%%%%%%%%%%%
%%%%%%%%%%%%%%%%%%%%%%%%
\begin{figure}
\centering
\includegraphics[angle=0,width=8cm]{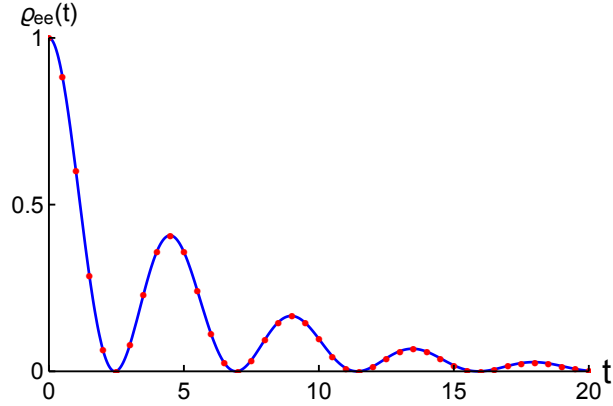}
\caption{\label{fig:fig2} The reduced dynamics of the spontaneous decay model with initial state $|\Phi(0)\rangle=|e\rangle$. The blue solid line is the exactly analytical result and the red circles are the numerical results obtained by HEOM. Other parameters are chosen as $\lambda=0.2$, $\gamma=5$ and $\omega_{0}=1$.}
\end{figure}
%%%%%%%%%%%%%%%%%%%%%%%%

\subsection{Self-adjoint coupling case}\label{sec:sec3a}

In the self-adjoint coupling operator case, i.e., $\hat{S}=\hat{S}^{\dag}$, one can combine $\mathbf{z}_{t}^{*}$ and $\mathbf{\tilde{z}}_{t}^{*}$ into a sum process $\mathbf{y}_{t}^{*}=\mathbf{z}_{t}^{*}+\mathbf{\tilde{z}}_{t}^{*}$, which can be easily demonstrated as a complex-valued Gaussian process as well. This sum complex-valued Gaussian process $\mathbf{y}_{t}^{*}$ satisfies $\mathbb{M}\{\mathbf{y}_{t}\}=\mathbb{M}\{\mathbf{y}_{t}^{*}\}=0$ and $\mathbb{M}\{\mathbf{y}_{t}\mathbf{y}_{\tau}^{*}\}=\xi(t-\tau)$, where $\xi(t)$ denotes the corresponding response function and is then given by
\begin{equation}\label{eq:eq35}
\begin{split}
\xi(t)\equiv&\alpha(t)+\tilde{\alpha}(t)\\
=&\int_{0}^{\infty}d\omega J(\omega)\Bigg{[}\coth\Bigg{(}\frac{\beta\omega}{2}\Bigg{)}\cos(\omega t)-i\sin(\omega t)\Bigg{]},
\end{split}
\end{equation}
which is also the well-known finite-temperature bath correlation function for the spin-boson model or the Caldeira-Leggett model. The non-Markovian quantum state diffusion equation in this case is given by
\begin{equation}\label{eq:eq36}
\partial_{t}|\Psi(\mathbf{y}_{t}^{*})\rangle=-i\hat{H}_{\mathrm{s}}|\Psi(\mathbf{y}_{t}^{*})\rangle+\hat{S}\mathbf{y}_{t}^{*}|\Psi(\mathbf{y}_{t}^{*})\rangle-\hat{S}\int_{0}^{t}d\tau\xi(t-\tau)\frac{\delta}{\delta \mathbf{y}_{\tau}^{*}}|\Psi(\mathbf{y}_{t}^{*})\rangle.
\end{equation}

For the SD scheme, in the self-adjoint coupling operator case, one can combine $w_{j1}(t)$ and $w_{j2}(t)$ into one new complex-valued Wiener process: $w_{j}(t)\equiv w_{j1}(t)+w_{j2}(t)$. Then, the stochastic differential equation given by Eq.~\ref{eq:eq22} reduces to
\begin{equation}\label{eq:eq37}
\begin{split}
d\hat{\tilde{\rho}}_{\mathrm{s}}(t)=&-i[\hat{H}_{\mathrm{s}}+\bar{g}(t)\hat{S},\hat{\tilde{\rho}}_{\mathrm{s}}(t)]dt-\frac{i}{2}[\hat{S},\hat{\tilde{\rho}}_{\mathrm{s}}(t)]dw_{1}+\frac{1}{2}\{\hat{S},\hat{\tilde{\rho}}_{\mathrm{s}}(t)\}dw_{2}^{*},
\end{split}
\end{equation}
where $\bar{g}(t)$ is the bath-induced random field
\begin{equation}\label{eq:eq38}
\begin{split}
\bar{g}(t)\equiv&\frac{\mathrm{tr}_{\mathrm{b}}[(\hat{B}+\hat{B}^{\dag})\hat{\rho}_{\mathrm{b}}(t)]}{\mathrm{tr}_{\mathrm{b}}[\hat{\rho}_{\mathrm{b}}(t)]}\\
=&\int_{0}^{t}d\tau\xi_{\mathrm{R}}(t-\tau)[\nu_{1}(\tau)-i\nu_{4}(\tau)]+\int_{0}^{t}d\tau\xi_{\mathrm{I}}(t-\tau)[\nu_{2}(\tau)+i\nu_{3}(\tau)],
\end{split}
\end{equation}
where $\nu_{i}(t)\equiv\sum_{j=1}^{2}\nu_{ij}(t)$ with $i=1,2,3,4$ are uncorrelated Gaussian white noises. The corresponding deterministic quantum master equation in this case is given by
\begin{equation}\label{eq:eq39}
\begin{split}
\frac{d}{dt}\hat{\varrho}_{\mathrm{s}}(t)=&-i[\hat{H}_{\mathrm{s}},\hat{\varrho}_{\mathrm{s}}(t)]-i\Bigg{[}\hat{S},\int_{0}^{t}d\tau\xi_{\mathrm{R}}(t-\tau)\mathcal{M}\Big{\{}\big{[}\nu_{1}(\tau)\hat{\tilde{\rho}}_{s}(t)-i\nu_{4}(\tau)\hat{\tilde{\rho}}_{s}(t)\big{]}\Big{\}}\Bigg{]}\\
&-i\Bigg{[}\hat{S},\int_{0}^{t}d\tau\xi_{\mathrm{I}}(t-\tau)\mathcal{M}\Big{\{}\big{[}\nu_{2}(\tau)\hat{\tilde{\rho}}_{s}(t)+i\nu_{3}(\tau)\hat{\tilde{\rho}}_{s}(t)\big{]}\Big{\}}\Bigg{]}.\\
\end{split}
\end{equation}
Then starting from Eq.~\ref{eq:eq36} or Eq.~\ref{eq:eq39}, one can construct the HEOM following the procedure shown in Sec.~\ref{sec:sec2}. It is necessary to point out that the number of the stochastic noises required in this case is reduced by half compared with the case $\hat{S}^{\dagger}\neq \hat{S}$, i.e., $\mathbf{z}_{t}^{*},\mathbf{\tilde{z}}_{t}^{*}\rightarrow\mathbf{y}_{t}^{*}$ and $ w_{j1}(t),w_{j2}(t)\rightarrow w_{j}(t)$. Next, we make a comparison between the numerical result obtained by the HEOM scheme and some exact results.

As an illustrative example, we consider an exactly solvable model, i.e., the pure dephasing model which can be described by Eq.~\ref{eq:eq1} with $\hat{H}_{\mathrm{s}}=\frac{1}{2}\omega_{0}\hat{\sigma}_{z}$ and $\hat{S}=\hat{S}^{\dagger}=\hat{\sigma}_{z}$. We assume the bath density spectral function is Ohmic spectrum with Drude cutoff, i.e., $J(\omega)=J_{\mathrm{O}}(\omega)$, where
\begin{equation}\label{eq:eq40}
J_{\mathrm{O}}(\omega)\equiv\frac{1}{\pi}\frac{2\chi\omega_{\mathrm{c}}\omega}{\omega^{2}+\omega_{\mathrm{c}}^{2}},
\end{equation}
with $\chi$ stands for the coupling strength between the quantum subsystem and the bath, parameter $\omega_{\mathrm{c}}$ is the cutoff frequency. In this case, the bath correlation function $\xi(t)$ is given by~\cite{39,58,59,60}
\begin{equation}\label{eq:eq41}
\begin{split}
\xi(t)=&\Bigg{[}\chi\omega_{\mathrm{c}}\cot\Bigg{(}\frac{\beta\omega_{\mathrm{c}}}{2}\Bigg{)}-i\chi\omega_{\mathrm{c}}\Bigg{]}e^{-\omega_{\mathrm{c}}t}+\frac{4\chi\omega_{\mathrm{c}}}{\beta}\sum_{\mathrm{n}=1}^{\infty}\frac{\vartheta_{\mathrm{n}}}{\vartheta_{\mathrm{n}}^{2}-\omega^{2}_{\mathrm{c}}}e^{-\vartheta_{\mathrm{n}} t},
\end{split}
\end{equation}
where $\vartheta_{\mathrm{n}}\equiv 2\mathrm{n}\pi/\beta$ denotes the $\mathrm{n}$-th Matsubara frequency. It appears that the bath correlation contains the sum-of-exponentials for the real and imaginary parts, respectively, which implies that four indexes $\mathrm{\mathbf{p}},\mathrm{\mathbf{q}},\mathrm{\mathbf{k}},\mathrm{\mathbf{l}}$ are needed. However, one can reexpressed Eq.~\ref{eq:eq41} as follows
\begin{equation}\label{eq:eq42}
\xi(t)\simeq\sum_{\mathrm{n}=0}^{\epsilon}\zeta_{\mathrm{n}}e^{-\kappa_{\mathrm{n}}t},
\end{equation}
where we have set an upper bound for the sum of exponentials. In fact, one can only consider the first few terms in the series of Eq.~\ref{eq:eq42}. This approximation is reliable when the bath temperature is not too low. The coefficients $\zeta_{\mathrm{n}}$ and $\kappa_{\mathrm{n}}$ are given by
\begin{equation*}
\zeta_{\mathrm{n}}\equiv\Bigg{[}\chi\omega_{c}\cot\Bigg{(}\frac{\beta\omega_{c}}{2}\Bigg{)}-i\chi\omega_{c}\Bigg{]}\delta_{\mathrm{n}0}+\frac{4\chi\omega_{c}}{\beta}\frac{\vartheta_{\mathrm{n}}}{\vartheta^{2}_{\mathrm{n}}-\omega^{2}_{c}}(1-\delta_{\mathrm{n}0}),
\end{equation*}
\begin{equation*}
\kappa_{\mathrm{n}}\equiv\omega_{\mathrm{c}}\delta_{\mathrm{n}0}+\vartheta_{\mathrm{n}}(1-\delta_{\mathrm{n}0}).
\end{equation*}
By doing so, there is only one sum-of-exponentials and mere two indexes $\mathrm{\mathbf{p}},\mathrm{\mathbf{k}}$ are required. This trick is applicable, because, in our HEOM formulation, all the fitting numbers $\{U_{n_{\mathrm{X}}}^{\mathrm{X}},V_{n_{\mathrm{X}}}^{\mathrm{X}},\tilde{U}_{\tilde{n}_{\mathrm{X}}}^{\mathrm{X}},\tilde{V}_{\tilde{n}_{\mathrm{X}}}^{\mathrm{X}}\}$ are allowed to be complex values.

Then, one can find that the indexes $\mathbf{u_{R}}=\{\zeta_{0},\zeta_{1},\zeta_{2},...,\zeta_{\epsilon}\}$, $\mathbf{v_{R}}=\{\kappa_{0},\kappa_{1},\kappa_{2},...,\kappa_{\epsilon}\}$, $\mathbf{p}=\{\mathrm{p}_{0},\mathrm{p}_{1},\mathrm{p}_{2},...,\mathrm{p}_{\epsilon}\}$ and $\mathbf{k}=\{\mathrm{k}_{0},\mathrm{k}_{1},\mathrm{k}_{2},...,\mathrm{k}_{\epsilon}\}$ reduces to $(\epsilon+1)$-dimensional vectors, and all the other indexes vanish. The hierarchy equations of the reduced quantum subsystem in this case are given by
\begin{equation}\label{eq:eq43}
\begin{split}
\partial_{t}\hat{\varrho}_{\mathbf{p}}^{\mathbf{k}}(t)=&-\frac{i}{2}\big{[}\omega_{0}\hat{\sigma}_{z},\hat{\varrho}_{\mathbf{p}}^{\mathbf{k}}(t)\big{]}-(\mathbf{p}\cdot \mathbf{v_{R}}+\mathbf{k}\cdot \mathbf{v_{R}^{*}})\hat{\varrho}_{\mathbf{p}}^{\mathbf{k}}(t)\\
&+\sum_{\mathrm{n}=0}^{\epsilon}\big{[}\mathrm{p}_{\mathrm{n}}\zeta_{\mathrm{n}}\hat{\sigma}_{z}\hat{\varrho}_{\mathbf{p}-\mathbf{e}_{\mathrm{n}}}^{\mathbf{k}}(t)+\mathrm{k}_{\mathrm{n}}\zeta_{\mathrm{n}}^{*}\hat{\varrho}_{\mathbf{p}}^{\mathbf{k}-\mathbf{e}_{\mathrm{n}}}(t)\hat{\sigma}_{z}\big{]}-\sum_{\mathrm{n}=0}^{\epsilon}\big{[}\hat{\sigma}_{z},\hat{\varrho}_{\mathbf{p}+\mathbf{e}_{\mathrm{n}}}^{\mathbf{k}}(t)-\hat{\varrho}_{\mathbf{p}}^{\mathbf{k}+\mathbf{e}_{\mathrm{n}}}(t)\big{]}.
\end{split}
\end{equation}
Note that this hierarchy equation is quite similar to that given in Ref.~\cite{33}. The slight difference originates from the choice of the auxiliary density operator, which does not effect the final numerical result. In Fig.~\ref{fig:fig1}, we display the dynamics of the quantum coherence $\langle\hat{\sigma}_{x}(t)\rangle\equiv \mathrm{tr_{s}}[\hat{\varrho}_{\mathrm{s}}(t)\hat{\sigma}_{x}]$ from the HEOM result and the exact result, and it is clear to see that they are in good agreement.

\subsection{Zero-temperature bath case}\label{sec:sec3b}

%%%%%%%%%%%%%%%%%%%%%%%%
\begin{figure}
\centering
\includegraphics[angle=0,width=8cm]{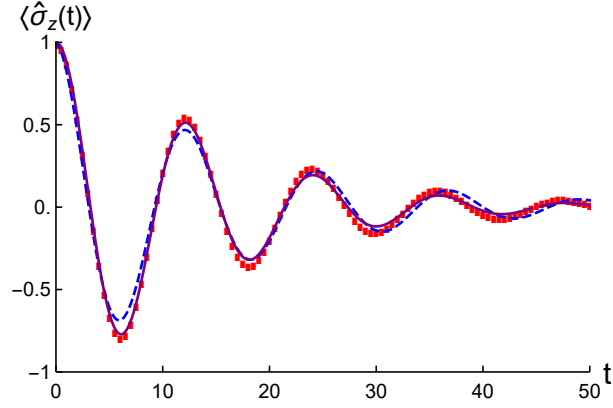}
\caption{\label{fig:fig3} The coherent dynamics of the population difference $\langle\hat{\sigma}_{z}(t)\rangle$ of the spin-boson model with initial state $|\Phi(0)\rangle=|e\rangle$ at zero temperature. The purple solid line is the numerical result from the HEOM method, the red rectangles denote the results obtained by the generalized Silbey-Harris transformation approach, and the blue dashed line represent the Wigner-Weisskopf result from Ref.\cite{radd1}. Other parameters are chosen as $\Delta=0.5$, $\lambda=0.5\gamma$, $\gamma=0.5$ and $\omega_{0}=\Delta$.}
\end{figure}
%%%%%%%%%%%%%%%%%%%%%%%%

In the zero-temperature bath case, i.e., when the bath is initially prepared in its Fock vacuum state, all the contributions arising from $\tilde{\alpha}(t)$ or $\grave{\alpha}(t)$ vanish, and one can easily obtain the corresponding HEOM by simply removing all the terms with $\sim$ in Eq.~\ref{eq:eq17}. For the zero-temperature case considered in this subsection, we assume the bath density spectral function has Lorentz form, namely, $J(\omega)=J_{\mathrm{L}}(\omega)$, where
\begin{equation}\label{eq:eq44}
J_{\mathrm{L}}(\omega)\equiv\frac{1}{2\pi}\frac{\gamma\lambda^{2}}{(\omega-\omega_{0})^{2}+\lambda^{2}},
\end{equation}
where $\lambda$ defines the spectral width of the coupling and $\gamma$ can be approximately interpreted as the system-bath coupling strength. For this Lorentz spectrum case, the bath correlation at zero temperature is given by~\cite{57}
\begin{equation}\label{eq:eq45}
\alpha_{\mathrm{L}}(t)=\frac{1}{2}\gamma\lambda\exp[-(\lambda+i\omega_{0}) t].
\end{equation}

 As the first illustrative example, we consider the famous spontaneous decay model which describes a two-level system that interacts with a vacuum; this model is given by Eq.~\ref{eq:eq1} with $\hat{H}_{\mathrm{s}}=\frac{1}{2}\omega_{0}\hat{\sigma}_{z}$ and $\hat{S}=\hat{\sigma}_{-}$. In this case, there is only one exponential which indicates that the indexes $\mathbf{u_{R}}=\{\gamma\lambda/2\}$, $\mathbf{v_{R}}=\{v\}=\{\lambda+i\omega_{0}\}$, $\mathbf{p}=\{\mathrm{p}\}$ and $\mathbf{k}=\{\mathrm{k}\}$ reduce to one-dimensional vectors, while all the other indexes vanish. Then, one can obtain the hierarchy equations of the reduced quantum subsystem as follows:
\begin{equation}\label{eq:eq46}
\begin{split}
\partial_{t}\hat{\varrho}_{\mathrm{p}}^{\mathrm{k}}(t)=&-\frac{i}{2}\big{[}\omega_{0}\hat{\sigma}_{z},\hat{\varrho}_{\mathrm{p}}^{\mathrm{k}}(t)\big{]}-\big{(}\mathrm{p}v+\mathrm{k}v^{*}\big{)}\hat{\varrho}_{\mathrm{p}}^{\mathrm{k}}(t)\\
&+\frac{1}{2}\gamma\lambda \big{[}\mathrm{p}\hat{\sigma}_{-}\hat{\varrho}_{\mathrm{p}-1}^{\mathrm{k}}(t)+\mathrm{k}\hat{\varrho}_{\mathrm{p}}^{\mathrm{k}-1}(t)\hat{\sigma}_{+}\big{]}-\big{[}\hat{\sigma}_{+},\hat{\varrho}_{\mathrm{p}+1}^{\mathrm{k}}(t)\big{]}+\big{[}\hat{\sigma}_{-},\hat{\varrho}_{\mathrm{p}}^{\mathrm{k}+1}(t)\big{]}.
\end{split}
\end{equation}
This equation is essentially the same with that given in Refs.~\cite{35,59}. In Fig.~\ref{fig:fig2}, we display the dynamics of $\varrho_{ee}(t)\equiv \mathrm{tr_{s}}[\hat{\varrho}_{\mathrm{s}}(t)|e\rangle\langle e|]$ obtained by the numerical HEOM method and the exactly analytical expression, it is clear to see that numerical results coincide perfectly with the exact analytical results.

The second example in this subsection is the spin-boson model without the rotating-wave approximation, i.e., $\hat{H}_{\mathrm{s}}=-\frac{1}{2}\Delta\hat{\sigma}_{x}$ and $\hat{S}=\frac{1}{2}\hat{\sigma}_{z}$. Generally, there is no exact dynamical result in this case, however, the spin-boson model can be approximately handled by making use of the generalized Silbey-Harris transformation~\cite{14,16,radd1,radd2} or the Wigner-Weisskopf approach~\cite{16,radd1,radd2}. Although these two methods neglect the higher-order terms of the system-bath coupling strength, it is still acceptable when the system-bath coupling is not too strong~\cite{16,59}. The HEOM of the spin-boson model can be easily derived by doing the following substitutions: $\omega_{0}\hat{\sigma}_{z}\rightarrow-\Delta\hat{\sigma}_{x}$ and $\hat{\sigma}_{\pm}\rightarrow\frac{1}{2}\hat{\sigma}_{z}$ in Eq.\ref{eq:eq46}. In Fig.~\ref{fig:fig3}, we display the numerical result from HEOM method, the results from the generalized Silbey-Harris transformation as well as the results from the Wigner-Weisskopf approach in a moderately strong-coupling regime. It is clear to see that the results from these three different methods are in qualitative agreement, which convinces us that our numerical HEOM scheme is reliable.

\section{Conclusion}\label{sec:sec4}

A stochastic formulation provides a very convenient way to describe the dynamical behaviour of a quantum open system, the reduced density matrix of the quantum subsystem can be obtained by simply taking the statistical average. Some stochastic dynamical approaches have been successfully applied to investigate the thermal distributions~\cite{61}, absorption or emission spectra~\cite{62} and energy transfer~\cite{63} problems in physical and chemical fields. The random noises invoked in the NMQSD and the SD schemes are complex-valued Gaussian noises because the bosonic bath considered in this paper is a Gaussian-type bath and the system-bath coupling is modeled at the linear hybridization level. It should be stressed that the stochastic formulation can be also extended to non-Gaussian-bath situations, such as the spin-bath system~\cite{26}. This makes the stochastic dynamical formulation a highly flexible tool in the field of open quantum system dynamics. Of course, the stochastic algorithm for a non-Gaussian bath is more difficult, because some higher order statistical terms, which are negligible in the Gaussian bath case, become nontrivial. It would be very interesting to generalize the NMSQD method and the SD approach to the non-Gaussian bath case.

In summary, we proposed a way to realize the HEOM for a generalized linear coupling quantum dissipative system in the frameworks of the NMQSD and the SD schemes. Although, a similar HEOM can be achieved by making use of the path integral influence functional approach~\cite{31,32,33,34,35,36,37}, we believe that any alternative method can help us to obtain more physical insights into the characters of quantum dissipative dynamics. Due to the fact that a stochastic simulation usually costs a great deal of computational resources and suffers from a numerically uncontrollable problem in the long-time regime, the main advantage of the HEOM is that one can perform a more reliable and deterministic simulation. When one investigates the dynamics of a quantum dissipative system, making such a change from the stochastic to the deterministic perspective is especially helpful, because, by doing so, the hybrid method combines the merits of the stochastic and the deterministic schemes. We demonstrated that the HEOMs obtained by these two different stochastic dynamical methods are identical. Generally speaking, the HEOM is beyond the Markovian approximation, the rotating-wave approximation, and the perturbative approximation. In other words, the HEOM contains all the physical information and fully determines the dissipative dynamics. In this sense, our result suggests that the NMQSD and the SD scheme are equivalent in terms of describing the dissipative dynamics under certain conditions. Moreover, we present three examples, i.e., the pure dephasing model, the spontaneous decay model and the famous spin-boson model beyond the rotating-wave approximation, to verify the feasibility of our method. The numerical results obtained by the HEOM approach are in good agreement with the results calculated by other approaches, which indicates our HEOM derived by stochastic formulations truly captures the dynamical behaviour of quantum dissipative systems. Finally, due to the generality of the qubit-oscillator model and the HEOM method, we expect our results to be of interest for a wide range of experimental applications in quantum dissipative systems.

\section{Acknowledgments}\label{sec:secack}

W. Wu wishes to thank Dr. Da-Wei Luo, Professor Jian-Qiang You and Professor Hai-Qing Lin for many useful discussions, W. Wu also acknowledges the fruitful communications with Professor Yun-An Yan, Professor Jiushu Shao and Professor Ting Yu during their short-term visits in Beijing Computational Science Research Center. This project is supported by the China Postdoctoral Science Foundation (Grant No.2017M610753), the NSFC (Grant No.11704025) and the NSAF (Grant No. U1530401).

\section{Appendix A: Stochastic Processes}\label{sec:secappa}

In the NMQSD method, the vacuum Fock states are specified by two sets of complex numbers $\{z_{\ell}\}$ and $\{\tilde{z}_{\ell}\}$, which are introduced to label the Bargmann coherent states and are regarded as random variables~\cite{20,21,22}. Due to the completeness relation of the Bargmann coherent state, one can easily find that $\mathbf{z}_{t}$ and $\mathbf{\tilde{z}}_{t}$ are two independent complex-valued colored Gaussian processes. In the NMQSD treatment, tracing out the degrees of freedom of the bath is equivalent to taking the statistical mean over the possible stochastic processes. The definition of the statistical mean $\mathbb{M}=\mathbb{M}_{\{z_{\ell}\},\{\tilde{z}_{\ell}\}}$ is given by
\begin{equation}\label{eq:eq47}
\begin{split}
\mathbb{M}\{F[\mathbf{z}_{t},\mathbf{\tilde{z}}_{t}]\}\equiv& \int_{-\infty}^{+\infty}\frac{d^{2}z_{1}}{\pi}e^{-|z_{1}|^{2}}\int_{-\infty}^{+\infty}\frac{d^{2}z_{2}}{\pi}e^{-|z_{2}|^{2}}...\int_{-\infty}^{+\infty}\frac{d^{2}\tilde{z}_{1}}{\pi}e^{-|\tilde{z}_{1}|^{2}}\int_{-\infty}^{+\infty}\frac{d^{2}\tilde{z}_{2}}{\pi}e^{-|\tilde{z}_{2}|^{2}}...\times \langle \mathbf{z}\mathbf{\tilde{z}}|F[\mathbf{z}_{t},\mathbf{\tilde{z}}_{t}]|\mathbf{z}\mathbf{\tilde{z}}\rangle.
\end{split}
\end{equation}
It is easy to check that $\mathbf{z}_{t}$ and $\mathbf{\tilde{z}}_{t}$ satisfy the statistical characteristic of complex-valued colored Gaussian processes, and their response functions are given by,
\begin{equation*}
\mathbb{M}\{\mathbf{z}_{t}\}=\mathbb{M}\{\mathbf{z}_{t}^{*}\}=0;~~~\mathbb{M}\{\mathbf{z}_{t}\mathbf{z}_{\tau}^{*}\}=\alpha(t-\tau),
\end{equation*}
\begin{equation*}
\mathbb{M}\{\mathbf{\tilde{z}}_{t}\}=\mathbb{M}\{\mathbf{\tilde{z}}_{t}^{*}\}=0;~~~\mathbb{M}\{\mathbf{\tilde{z}}_{t}\mathbf{\tilde{z}}_{\tau}^{*}\}=\tilde{\alpha}(t-\tau),
\end{equation*}
where $\alpha(t)$ and $\tilde{\alpha}(t)$ are corresponding modified bath correlation functions defined by Eq.~\ref{eq:eq3} and Eq.~\ref{eq:eq4}, respectively. In this sense, the effect of the bath on the quantum subsystem is fully characterized by the statistical characteristics of stochastic processes $\mathbf{z}_{t}$ and $\mathbf{\tilde{z}}_{t}$.

For the SD scheme proposed by Shao \emph{et al}, the decoupling is achieved by the It$\hat{\mathrm{o}}$ calculus. The It$\hat{\mathrm{o}}$ calculus is concerned with a Wiener process, say, $\varpi(t)\equiv\int_{0}^{t}d\tau\mu(\tau)$ where $\mu(t)$ is a Gaussian white noise with zero mean and delta function correlation, i.e., $\mathcal{M}\{\mu(t)\}=0$ and $\mathcal{M}\{\mu(t)\mu(\tau)\}=\delta(t-\tau)$. Taking a uniform discretization of a given time domain $[0,t]$ and assuming that each time interval $\Delta t=\Delta t_{i}=t_{i}-t_{i-1}$ is infinitesimal, one can roughly regard the white noise $\mu(t)$ as a series of independent random numbers at time slices. Thus the distribution or the weight function for any $\mu_{i}=\mu(t_{i})$ is given by~\cite{25}
\begin{equation}\label{eq:eq48}
W(\mu_{i})=\lim_{\Delta t\rightarrow 0}\sqrt{\frac{\Delta t}{2\pi}}\exp\Bigg{(}-\frac{\Delta t}{2}\mu_{i}^{2}\Bigg{)}.
\end{equation}
For an arbitrary functional $\mathcal{F}(\mu_{1},\mu_{2},\mu_{3},...)$ in discrete time representation, the statistical average of $\mathcal{F}(\mu_{1},\mu_{2},\mu_{3},...)$ is defined by
\begin{equation}\label{eq:eq49}
\mathcal{M}\{\mathcal{F}(\mu_{1},\mu_{2},\mu_{3},...)\}\equiv\Bigg{[}\prod_{i}\int_{-\infty}^{+\infty}d\mu_{i}W({\mu_{i}})\Bigg{]}\mathcal{F}(\mu_{1},\mu_{2},\mu_{3},...).
\end{equation}
Though a Wiener process can be viewed as a special Gaussian process, in this paper, we prefer to use a different notation $\mathcal{M}$ rather than $\mathbb{M}$ to signify the statistical average.

For a conventional Wiener process, the It$\hat{\mathrm{o}}$ calculus states that $(d\varpi)^{2}=dt$ and $d\varpi d\varpi'=0$ where $\varpi'(t)$ is another Wiener process~\cite{64}. However, in the SD scheme, $w_{1j}(t)$ and $w_{2j}(t)$ with $j=1,2$ are complex-valued Wiener processes. As a result, a straightforwardly generalization of the It$\hat{\mathrm{o}}$ calculus says that the four complex-valued Wiener processes, which are involved in the SD approach, satisfy the following relations~\cite{24,25,27}
\begin{equation}\label{eq:eq50}
dw_{1j}(t)dw_{1j'}(t)=dw_{2j}(t)dw_{2j'}(t)=dw_{1j}^{*}(t)dw_{1j'}^{*}(t)=dw_{2j}^{*}(t)dw_{2j'}^{*}(t)=0,
\end{equation}
and
\begin{equation}\label{eq:eq51}
dw_{1j}(t)dw_{1j'}^{*}(t)=dw_{2j}(t)dw_{2j'}^{*}(t)=2\delta_{jj'}dt.
\end{equation}
Together with the non-anticipating function's property, one can find that
\begin{equation*}
\begin{split}
d\mathcal{M}\{\hat{\rho}_{\mathrm{s}}(t)\hat{\rho}_{\mathrm{b}}(t)\}=&\mathcal{M}\{[d\hat{\rho}_{\mathrm{s}}(t)]\hat{\rho}_{\mathrm{b}}(t)+\hat{\rho}_{\mathrm{s}}(t)[d\hat{\rho}_{\mathrm{b}}(t)]+d\hat{\rho}_{\mathrm{s}}(t)d\hat{\rho}_{\mathrm{b}}(t)\}\\
=&\mathcal{M}\{-i[\hat{H}_{\mathrm{s}},\hat{\rho}_{\mathrm{s}}(t)]\hat{\rho}_{\mathrm{b}}(t)dt\}+\mathcal{M}\{-i[\hat{H}_{\mathrm{b}},\hat{\rho}_{\mathrm{b}}(t)]\hat{\rho}_{\mathrm{s}}(t)dt\}\\
&-\frac{i}{4}\mathcal{M}\{[\hat{S},\hat{\rho}_{\mathrm{s}}(t)]\{\hat{B}^{\dag},\hat{\rho}_{\mathrm{b}}(t)\}dw_{11}dw_{11}^{*}\}-\frac{i}{4}\mathcal{M}\{[\hat{S}^{\dag},\hat{\rho}_{\mathrm{s}}(t)]\{\hat{B},\hat{\rho}_{\mathrm{b}}(t)\}dw_{12}dw_{12}^{*}\}\\
&-\frac{i}{4}\mathcal{M}\{\{\hat{S},\hat{\rho}_{\mathrm{s}}(t)\}[\hat{B}^{\dag},\hat{\rho}_{\mathrm{b}}(t)]dw_{21}dw_{21}^{*}\}-\frac{i}{4}\mathcal{M}\{\{\hat{S}^{\dag},\hat{\rho}_{\mathrm{s}}(t)\}[\hat{B},\hat{\rho}_{\mathrm{b}}(t)]dw_{22}dw_{22}^{*}\}\\
=&-i[\hat{H}_{\mathrm{s}}+\hat{H}_{\mathrm{b}}+\hat{S}\hat{B}^{\dagger}+\hat{S}^{\dagger}\hat{B},\mathcal{M}\{\hat{\rho}_{\mathrm{s}}(t)\hat{\rho}_{\mathrm{b}}(t)\}]dt\\
=&-i[\hat{H},\mathcal{M}\{\hat{\rho}_{\mathrm{s}}(t)\hat{\rho}_{\mathrm{b}}(t)\}]dt.\\
\Rightarrow&\frac{d}{dt}\hat{\varrho}_{\mathrm{sb}}(t)=-i[\hat{H},\hat{\varrho}_{\mathrm{sb}}(t)].
\end{split}
\end{equation*}
This result demonstrates that $\mathcal{M}\{\hat{\rho}_{\mathrm{s}}(t)\hat{\rho}_{\mathrm{b}}(t)\}$ indeed satisfies the quantum Liouville equation given by Eq.~\ref{eq:eq18} under our definition of the It$\hat{\mathrm{o}}$ calculus given by Eq.~\ref{eq:eq50} and Eq.~\ref{eq:eq51}. In other words, the SD scheme is a self-consistent formalism.
The same statistical characteristics of the complex-valued Wiener processes (namely, Eq.~\ref{eq:eq50} and Eq.~\ref{eq:eq51}) in stochastic dynamical description are also chosen in some previous studies~\cite{26,add2}.

\section{Appendix B: Decomposition of bath correlation functions}\label{sec:secappb}

There are many different strategies to approximately decompose the modified bath correlation functions as a finite sum of exponentials, such as the Matsubara decomposition~\cite{add3}, the famous Meier-Tannor decomposition and its extensions~\cite{44,45}, the continued fraction expansion~\cite{47}, and the Pad$\mathrm{\acute{e}}$ decomposition~\cite{46}. Different decomposition schemes are mathematically equivalent, but may have different numerical performances. It has been reported that the numerical performance of the Pad$\mathrm{\acute{e}}$ decomposition is much better (about an order of magnitude faster) than the conventional formalism based on the Matsubara decomposition scheme~\cite{65}. All the mentioned schemes are based on Cauchy's residue theorem which implies that we need to choose a suitable integration contour.

Here, we will only sketch the basic idea following the detailed exposition in Ref.~\cite{44,45}. Considering the fact that that $\acute{\alpha}(t)=\alpha^{*}(t)-\tilde{\alpha}(t)$ and $\grave{\alpha}(t)=\alpha^{*}(t)+\tilde{\alpha}(t)$, it is more convenient to decompose $\acute{\alpha}(t)$ and $\grave{\alpha}(t)$ instead of $\alpha(t)$ and $\tilde{\alpha}(t)$, because the hyperbolic cotangent is anti-symmetric in $\omega$ which would be very helpful to extend the integral boundary in Eq.~\ref{eq:eq26} and design a correct integration contour. To this aim, we introduce an anti-symmetric continuation bath density spectrum~\cite{45}
\begin{equation}\label{eq:eq52}
\mathcal{J}(\omega)=\begin{cases}
J(\omega),~~~\omega\geq 0;\\[6pt]
-J(-\omega),~~~\omega<0.
\end{cases}
\end{equation}
Then one can expand the integral upper boundary in Eq.~\ref{eq:eq26} to the whole real axis, i.e.,
\begin{equation*}
\begin{split}
\grave{\alpha}_{\mathrm{R}}(t)=&\frac{1}{2}\int_{-\infty}^{+\infty}d\omega \mathcal{J}(\omega)\coth\Bigg{(}\frac{\beta\omega}{2}\Bigg{)}\cos(\omega t)\\
=&\frac{1}{2}\int_{-\infty}^{+\infty}d\omega \mathcal{J}(\omega)\coth\Bigg{(}\frac{\beta\omega}{2}\Bigg{)}e^{i\omega t}.
\end{split}
\end{equation*}
Let $\varsigma$ denotes those poles of $\mathcal{J}(\omega)\coth(\beta\omega/2)$ that lie within the integration contour, then one has that
\begin{equation*}
\grave{\alpha}_{\mathrm{R}}(t)=i\pi\sum_{\varsigma}\mathrm{Res}_{\varsigma}\Bigg{[}\mathcal{J}(\omega)\coth\Bigg{(}\frac{\beta\omega}{2}\Bigg{)}\Bigg{]}e^{i\varsigma t},
\end{equation*}
where we have used the celebrated Cauchy's residue theorem. Assuming that the poles $\varsigma_{1}$ of the fitting spectrum $\mathcal{J}(\omega)$ do not coincide with the poles $\varsigma_{2}$ of the hyperbolic cotangent, one can obtain
\begin{equation*}
\grave{\alpha}_{\mathrm{R}}(t)=i\pi\sum_{\varsigma_{1}}\mathrm{Res}_{\varsigma_{1}}[\mathcal{J}(\omega)]\coth\Bigg{(}\frac{\beta\varsigma_{1}}{2}\Bigg{)}e^{i\varsigma_{1} t}+i\pi\sum_{\varsigma_{2}}\mathrm{Res}_{\varsigma_{2}}\Bigg{[}\coth\Bigg{(}\frac{\beta\omega}{2}\Bigg{)}\Bigg{]}\mathcal{J}(\varsigma_{2})e^{i\varsigma_{2} t}.
\end{equation*}

It is necessary to point out that the number of poles may be infinite in certain situations, say the innumerable poles arise from the Matsubara expansion of the hyperbolic cotangent or the Bose-Einstein function $(1-e^{-\beta\omega})-1$, one needs to truncate the summation over the terms for practical purposes.

Making use of a similar method, one can also approximately express $\grave{\alpha}_{\mathrm{I}}(t)$, $\acute{\alpha}_{\mathrm{R}}(t)$ and $\acute{\alpha}_{\mathrm{I}}(t)$ as a sum of exponentials. As demonstrated in Ref.~\cite{45}, such a decomposition scheme can be used to handle the Ohmic as well as the super-Ohmic spectral densities at finite temperatures by using appropriate fitting routines. We also like to point out that some more generalized methods to find a good fitting of the bath correlation function have been proposed in recent years, in these methods, the fitting parameters $\{U_{n_{\mathrm{X}}}^{\mathrm{X}},V_{n_{\mathrm{X}}}^{\mathrm{X}},\tilde{U}_{\tilde{n}_{\mathrm{X}}}^{\mathrm{X}},\tilde{V}_{\tilde{n}_{\mathrm{X}}}^{\mathrm{X}}\}$ are allowed to be time-dependent~\cite{66,67,68}. For example, in Refs.~\cite{67,68}, the authors introduced a complete set of orthonormal basis vectors to expand the bath correlation function and obtained a highly-accurate fitting function with very few terms. They also used this approach to investigate the delocalized-localized quantum phase transition in a sub-Ohmic spin-boson model at zero temperature~\cite{68}.

\section{Appendix C: Extended Furutsu-Novikov Theorem}\label{sec:secappc}

In this appendix, we shall prove Eqs.~\ref{eq:eq13}-\ref{eq:eq16} by making use of the extended Furutsu-Novikov theorem for complex-valued Gaussian noises which has been proven in several previous Refs.~\cite{22,29,50,69}. The extended Furutsu-Novikov theorem states that the ensemble average of the functional $\mathfrak{F}_{t}\equiv\mathfrak{F}(\mathbf{z}_{t},\mathbf{\tilde{z}}_{t})=|\Psi_{\mathbf{p},\mathbf{q},\tilde{\mathbf{p}},\tilde{\mathbf{q}}}\rangle\langle\Psi_{\mathbf{k},\mathbf{l},\tilde{\mathbf{k}},\tilde{\mathbf{l}}}|$, which depends on any complexe-valued Gaussian noise $\eta_{t}=\mathbf{z}_{t}$ or $\mathbf{\tilde{z}}_{t}$, equals
\begin{equation}\label{eq:eq53}
\mathbb{M}\{\mathfrak{F}_{t}\eta_{t}\}=\int_{0}^{t} d\tau\mathbb{M}\{\eta_{t}\eta^{*}_{\tau}\}\mathbb{M}\Bigg{\{}\frac{\overrightarrow{\delta}}{\delta\eta^{*}_{\tau}} \mathfrak{F}_{t}\Bigg{\}},
\end{equation}
\begin{equation}\label{eq:eq54}
\mathbb{M}\{\eta^{*}_{t}\mathfrak{F}_{t}\}=\int_{0}^{t} d\tau\mathbb{M}\{\eta^{*}_{t}\eta_{\tau}\}\mathbb{M}\Bigg{\{} \mathfrak{F}_{t}\frac{\overleftarrow{\delta}}{\delta\eta_{\tau}}\Bigg{\}}.
\end{equation}
If $\eta_{t}$ is a Gaussian white noise, the corresponding response function reduces to the Dirac-$\delta$ function, and one can find that Eq.~\ref{eq:eq53} and Eq.~\ref{eq:eq54} recover the conventional Furutsu-Novikov theorem which is widely used in Refs.~\cite{23,24,25,27,51}.

Making use of Eq.~\ref{eq:eq53} and Eq.~\ref{eq:eq54}, one can find that
\begin{equation*}
\begin{split}
\mathbb{M}\{|\Psi_{\mathbf{p},\mathbf{q},\tilde{\mathbf{p}},\tilde{\mathbf{q}}}\rangle\langle\Psi_{\mathbf{k},\mathbf{l},\tilde{\mathbf{k}},\tilde{\mathbf{l}}}|\mathbf{z}_{t}\}=&\int_{0}^{t}d\tau\mathbb{M}\{\mathbf{z}_{t}\mathbf{z}_{\tau}^{*}\}\mathbb{M}\Bigg{\{}\frac{\overrightarrow{\delta}}{\delta \mathbf{z}_{\tau}^{*}}|\Psi_{\mathbf{p},\mathbf{q},\tilde{\mathbf{p}},\tilde{\mathbf{q}}}\rangle\langle\Psi_{\mathbf{k},\mathbf{l},\tilde{\mathbf{k}},\tilde{\mathbf{l}}}|\Bigg{\}}=\mathbb{M}\Bigg{\{}\int_{0}^{t}d\tau\alpha(t-\tau)\frac{\overrightarrow{\delta}}{\delta \mathbf{z}_{\tau}^{*}}|\Psi_{\mathbf{p},\mathbf{q},\tilde{\mathbf{p}},\tilde{\mathbf{q}}}\rangle\langle\Psi_{\mathbf{k},\mathbf{l},\tilde{\mathbf{k}},\tilde{\mathbf{l}}}|\Bigg{\}}\\
=&\mathbb{M}\Bigg{\{}\int_{0}^{t}d\tau\alpha_{\mathrm{R}}(t-\tau)\frac{\overrightarrow{\delta}}{\delta \mathbf{z}_{\tau}^{*}}|\Psi_{\mathbf{p},\mathbf{q},\tilde{\mathbf{p}},\tilde{\mathbf{q}}}\rangle\langle\Psi_{\mathbf{k},\mathbf{l},\tilde{\mathbf{k}},\tilde{\mathbf{l}}}|\Bigg{\}}+\mathbb{M}\Bigg{\{}i\int_{0}^{t}d\tau\alpha_{\mathrm{I}}(t-\tau)\frac{\overrightarrow{\delta}}{\delta \mathbf{z}_{\tau}^{*}}|\Psi_{\mathbf{p},\mathbf{q},\tilde{\mathbf{p}},\tilde{\mathbf{q}}}\rangle\langle\Psi_{\mathbf{k},\mathbf{l},\tilde{\mathbf{k}},\tilde{\mathbf{l}}}|\Bigg{\}}\\
=&\mathbb{M}\Big{\{}\sum_{n_{\mathrm{R}}}\mathfrak{\hat{D}}_{n_{\mathrm{R}}}^{\mathrm{R}}|\Psi_{\mathbf{p},\mathbf{q},\tilde{\mathbf{p}},\tilde{\mathbf{q}}}\rangle\langle\Psi_{\mathbf{k},\mathbf{l},\tilde{\mathbf{k}},\tilde{\mathbf{l}}}|\Big{\}}+\mathbb{M}\Big{\{}\sum_{n_{\mathrm{I}}}\mathfrak{\hat{D}}_{n_{\mathrm{I}}}^{\mathrm{I}}|\Psi_{\mathbf{p},\mathbf{q},\tilde{\mathbf{p}},\tilde{\mathbf{q}}}\rangle\langle\Psi_{\mathbf{k},\mathbf{l},\tilde{\mathbf{k}},\tilde{\mathbf{l}}}|\Big{\}}\\
=&\sum_{n_{\mathrm{R}}}\hat{\varrho}_{\mathbf{p}+\mathbf{e}_{n_{\mathrm{R}}},\mathbf{q},\mathbf{\tilde{p}},\mathbf{\tilde{q}}}^{\mathbf{k},\mathbf{l},\mathbf{\tilde{k}},\mathbf{\tilde{l}}}+\sum_{n_{\mathrm{I}}}\hat{\varrho}_{\mathbf{p},\mathbf{q}+\mathbf{e}_{n_{\mathrm{I}}},\mathbf{\tilde{p}},\mathbf{\tilde{q}}}^{\mathbf{k},\mathbf{l},\mathbf{\tilde{k}},\mathbf{\tilde{l}}}.
\end{split}
\end{equation*}
Using the same method, one can also prove Eq.~\ref{eq:eq14}, and
\begin{equation*}
\begin{split}
\mathbb{M}\{\mathbf{z}_{t}^{*}|\Psi_{\mathbf{p},\mathbf{q},\tilde{\mathbf{p}},\tilde{\mathbf{q}}}\rangle\langle\Psi_{\mathbf{k},\mathbf{l},\tilde{\mathbf{k}},\tilde{\mathbf{l}}}|\}=&\int_{0}^{t}d\tau\mathbb{M}\{\mathbf{z}_{t}^{*}\mathbf{z}_{\tau}\}\mathbb{M}\Bigg{\{}|\Psi_{\mathbf{p},\mathbf{q},\tilde{\mathbf{p}},\tilde{\mathbf{q}}}\rangle\langle\Psi_{\mathbf{k},\mathbf{l},\tilde{\mathbf{k}},\tilde{\mathbf{l}}}|\frac{\overleftarrow{\delta}}{\delta \mathbf{z}_{\tau}}\Bigg{\}}=\mathbb{M}\Bigg{\{}\int_{0}^{t}d\tau\tilde{\alpha}^{*}(t-\tau)|\Psi_{\mathbf{p},\mathbf{q},\tilde{\mathbf{p}},\tilde{\mathbf{q}}}\rangle\langle\Psi_{\mathbf{k},\mathbf{l},\tilde{\mathbf{k}},\tilde{\mathbf{l}}}|\frac{\overleftarrow{\delta}}{\delta \mathbf{z}_{\tau}}\Bigg{\}}\\
=&\mathbb{M}\Bigg{\{}\int_{0}^{t}d\tau\tilde{\alpha}_{\mathrm{R}}(t-\tau)|\Psi_{\mathbf{p},\mathbf{q},\tilde{\mathbf{p}},\tilde{\mathbf{q}}}\rangle\langle\Psi_{\mathbf{k},\mathbf{l},\tilde{\mathbf{k}},\tilde{\mathbf{l}}}|\frac{\overleftarrow{\delta}}{\delta \mathbf{z}_{\tau}}\Bigg{\}}+\mathbb{M}\Bigg{\{}-i\int_{0}^{t}d\tau\tilde{\alpha}_{\mathrm{I}}(t-\tau)|\Psi_{\mathbf{p},\mathbf{q},\tilde{\mathbf{p}},\tilde{\mathbf{q}}}\rangle\langle\Psi_{\mathbf{k},\mathbf{l},\tilde{\mathbf{k}},\tilde{\mathbf{l}}}|\frac{\overleftarrow{\delta}}{\delta \mathbf{z}_{\tau}}\Bigg{\}}\\
=&\mathbb{M}\Big{\{}\sum_{n_{\mathrm{R}}}|\Psi_{\mathbf{p},\mathbf{q},\tilde{\mathbf{p}},\tilde{\mathbf{q}}}\rangle\langle\Psi_{\mathbf{k},\mathbf{l},\tilde{\mathbf{k}},\tilde{\mathbf{l}}}|\mathfrak{\hat{D}}_{n_{\mathrm{R}}}^{\mathrm{R}\dag}\Big{\}}+\mathbb{M}\Big{\{}\sum_{n_{\mathrm{I}}}|\Psi_{\mathbf{p},\mathbf{q},\tilde{\mathbf{p}},\tilde{\mathbf{q}}}\rangle\langle\Psi_{\mathbf{k},\mathbf{l},\tilde{\mathbf{k}},\tilde{\mathbf{l}}}|\mathfrak{\hat{D}}_{n_{\mathrm{I}}}^{\mathrm{I}\dag}\Big{\}}\\
=&\sum_{n_{\mathrm{R}}}\hat{\varrho}_{\mathbf{p},\mathbf{q},\mathbf{\tilde{p}},\mathbf{\tilde{q}}}^{\mathbf{k}+\mathbf{e}_{n_{\mathrm{R}}},\mathbf{l},\mathbf{\tilde{k}},\mathbf{\tilde{l}}}+\sum_{n_{\mathrm{I}}}\hat{\varrho}_{\mathbf{p},\mathbf{q},\mathbf{\tilde{p}},\mathbf{\tilde{q}}}^{\mathbf{k},\mathbf{l}+\mathbf{e}_{n_{\mathrm{I}}},\mathbf{\tilde{k}},\mathbf{\tilde{l}}}.
\end{split}
\end{equation*}
Employing the same procedure, one can demonstrate Eq.~\ref{eq:eq16} as well.

\end{document}